\documentclass[12pt,a4paper]{article}

\usepackage{amsmath}
\usepackage{graphicx}
\usepackage{enumerate}
\usepackage{url}
\usepackage{float}
\usepackage{ragged2e}
\usepackage{amsthm}
\usepackage{mathtools}
\usepackage{amsfonts}
\usepackage{amssymb}
\usepackage{hyperref}  
\usepackage[subtle]{savetrees} 
\usepackage{caption}
\usepackage{subcaption}
\usepackage{apptools}
\usepackage{xcolor}
\usepackage{xr}
\usepackage[margin=1in]{geometry}
\DeclareRobustCommand{\E}{\mathbb{E}}
\DeclareRobustCommand{\N}{\mathbb{N}}
\DeclareRobustCommand{\Q}{\mathbb{Q}}
\DeclareRobustCommand{\H}{\mathbb{H}}
\DeclareRobustCommand{\L}{\mathbb{L}}
\DeclareRobustCommand{\R}{\mathbb{R}}
\DeclareRobustCommand{\P}{\mathbb{P}}
\DeclareRobustCommand{\DS}{\textsc{ds}}
\DeclareRobustCommand{\EXP}{\textsc{exp}}
\DeclareRobustCommand{\OBS}{\textsc{obs}}

\DeclareRobustCommand{\cA}{\mathcal{A}}
\DeclareRobustCommand{\cD}{\mathcal{D}}
\DeclareRobustCommand{\cB}{\mathcal{B}}
\DeclareRobustCommand{\cL}{\mathcal{L}}
\DeclareRobustCommand{\cC}{\mathcal{C}}
\DeclareRobustCommand{\cY}{\mathcal{Y}}
\DeclareRobustCommand{\cS}{\mathcal{S}}
\DeclareRobustCommand{\op}{\text{op}}
\DeclareRobustCommand{\cX}{\mathcal{X}}
\DeclareRobustCommand{\Span}{\text{span}}
\DeclareRobustCommand{\tr}{\text{trace}}
\DeclareMathOperator*{\argmin}{\arg\!\min}
\DeclareMathOperator*{\argmax}{\arg\!\max}
\DeclareRobustCommand{\d}{\mathrm{d}}

\newcommand{\indep}{\raisebox{0.05em}{\rotatebox[origin=c]{90}{$\models$}}}

\newcommand*{\centernot}{%
  \mathpalette\@centernot
}
\def\@centernot#1#2{%
  \mathrel{%
    \rlap{%
      \settowidth\dimen@{$\m@th#1{#2}$}%
      \kern.5\dimen@
      \settowidth\dimen@{$\m@th#1=$}%
      \kern-.5\dimen@
      $\m@th#1\not$%
    }%
    {#2}%
  }%
}

\theoremstyle{definition}

\newtheorem{theorem}{Theorem}
\newtheorem{assumption}{Assumption}
\newtheorem{algorithm}{Algorithm}
\newtheorem{proposition}{Proposition}
\newtheorem{definition}{Definition}
\newtheorem{remark}{Remark}
\newtheorem{corollary}{Corollary}
\newtheorem{lemma}{Lemma}
\newtheorem{example}{Example}

\AtAppendix{\counterwithin{theorem}{section}}
\AtAppendix{\counterwithin{assumption}{section}}
\AtAppendix{\counterwithin{algorithm}{section}}
\AtAppendix{\counterwithin{proposition}{section}}
\AtAppendix{\counterwithin{definition}{section}}
\AtAppendix{\counterwithin{remark}{section}}
\AtAppendix{\counterwithin{corollary}{section}}
\AtAppendix{\counterwithin{lemma}{section}}
\AtAppendix{\counterwithin{example}{section}}

\begin{document}

\def\spacingset#1{\renewcommand{\baselinestretch}%
{#1}\small\normalsize} \spacingset{1}


  \title{Kernel methods for \\
long term dose response curves}
  \author{
  Rahul Singh\thanks{Email: \url{rahul_singh@fas.harvard.edu}. Address: Littauer Center 123, 1805 Cambridge Street, Cambridge, MA 02138. We thank Raj Chetty for helpful discussions. Leonard Mushunje provided excellent research assistance.
  } \\
  Harvard University
  \and 
    Hannah Zhou \\
      Harvard University
  }
  \date{Original draft: January 2022. This draft: December 2024.}
\maketitle
%
\begin{abstract}
A core challenge in causal inference is how to extrapolate long term effects, of possibly continuous actions, from short term experimental data. 
It arises in artificial intelligence: the long term consequences of continuous actions may be of interest, yet only short term rewards may be collected in exploration. 
For this estimand, called the long term dose response curve, we propose a simple nonparametric estimator based on kernel ridge regression.
By embedding the distribution of the short term experimental data with kernels, we derive interpretable weights for extrapolating long term effects.
Our method allows actions, short term rewards, and long term rewards to be continuous in general spaces. It also allows for nonlinearity and heterogeneity in the link between short term effects and long term effects.
We prove uniform consistency, with nonasymptotic error bounds reflecting the effective dimension of the data.
As an application, we estimate the long term dose response curve of Project STAR, a social program which randomly assigned students to various class sizes.
We extend our results to long term counterfactual distributions, proving weak convergence.
\end{abstract}

\noindent%

\vfill

\newpage

\spacingset{1.9} 

\section{Introduction and related work}\label{sec:intro}

Experimental data have randomized actions, yet they are typically limited in duration due to the expense of conducting an experiment. 
From an artificial intelligence perspective, the long term consequence of an action may be of interest, yet only short term rewards may be collected in exploration.
From a program evaluation perspective, the long term effect of an action is what often matters, yet only short term outcomes may be collected in experimental surveys.

The goal of long term causal inference is to properly combine short term experimental data, which contain randomized actions, with long term observational data, which lack randomized actions \cite{prentice1989surrogate,freedman1992statistical,begg2000use,frangakis2002principal}.
Our research question is how to extrapolate long term effects of continuous actions, allowing nonlinearity and heterogeneity in the link between the short term and the long term.

We study nonparametric causal functions of a continuous action $D$,  which may refer to intervention intensities, medical dosages, or program lengths. For example, Project STAR randomly assigned kindergarten students to various class sizes $D$, ranging from 12 to 28 students. The short term test scores $S$ of these students were collected, however their long term earnings $Y$ were not. Separately, researchers may have access to test scores $S$ and earnings $Y$ for students in another school district.
Several social scientists \cite{krueger2001effect,chetty2011does} have asked: how much would the average student benefit, in the long term, from enrolling in a certain kindergarten class size? This quantity is an example of a long term dose response $\E\{Y^{(d)}\}$, where the class size is the dose, and the expected long term benefit is the response.

The difficulty in estimating long term dose response curves is the complex nonlinearity and heterogeneity in the link between short term response curve $\E\{S^{(d)}\}$ and long term response curve $\E\{Y^{(d)}\}$. For example, we would like to allow for the link between counterfactual test scores and counterfactual earnings to be nonlinear, and for students with different baseline characteristics to have different links. The identification of long term dose response curves is well known \cite{rosenman2018propensity,athey2020estimating,athey2020combining,rosenman2020combining, kallus2020role}, however it appears that no nonlinear, nonparametric estimators have been previously proposed for the response curve, highlighting the difficulty of extrapolating the long term effects of continuous actions.

Our primary contribution is to propose and analyze what appears to be the first nonlinear, nonparametric estimator for the long term dose response curve. 
We adapt kernel ridge regression, a classic statistical algorithm \cite{wahba1990spline}, to this learning problem.
By embedding the distribution of the short term experimental data using kernels, we derive interpretable weights for extrapolating long term effects from short term effects.
The final estimator has a simple closed form solution, while preserving nonlinearity and heterogeneity in the link between the short term and long term.
As an extension, we study long term counterfactual distributions in Appendix~\ref{sec:distribution}.

We use the statistical framework of the reproducing kernel Hilbert space (RKHS) \cite{steinwart2008support,berlinet2011reproducing} to overcome the technical issues that arise when estimating long term dose response curves. 
The identifying formula for the long term dose response curve is generally unbounded when actions are continuous \cite{van1991differentiable,newey1994asymptotic}, but we prove that it is bounded in the RKHS. 
Therefore, our proposed kernel embedding of short term experimental data exists. To adjust for how experiments may be limited in not only duration but also scope, we further define and characterize a ``fused'' kernel embedding.

Using this embedding, we prove the estimator is uniformly consistent across actions, with finite sample rates that combine well known minimax rates for kernel ridge regression.
The rates do not directly depend on the ambient dimension of the data, but rather the effective dimension and smoothness of the data with respect to an approximating basis \cite{caponnetto2007optimal,fischer2017sobolev}. The approximating basis is the spectrum of a kernel chosen by the researcher. See Section~\ref{sec:rkhs} for interpretation and comparison to Sobolev rates.

We illustrate the practicality of our approach by estimating the long term dose response of Project STAR, modelling class size as a continuous action. By allowing for continuous actions and heterogeneous links, our long term dose response estimate suggests that the effects of class size are nonlinear. Using short term experimental data and long term observational data, our method measures similar long term effects as an oracle method that has access to long term experimental data.

Several works provide strong semiparametric guarantees for the long term effects of binary actions \cite{kallus2020role,chen2021semiparametric,meza2021nested}. 
By contrast, we prove what appear to be the first nonparametric guarantees for the nonlinear, long term dose response curves of continuous actions.
Whereas binary actions imply boundedness of the identifying formula, continuous actions imply unboundedness in general; see Section~\ref{sec:problem}. Continuous actions introduce technical issues that we overcome with RKHS techniques.

We complement a literature that incorporates RKHS techniques into the estimation of other causal functions. 
For experimental data with joint observations of the randomized actions and final rewards, previous works analyze heterogeneous effects \cite{nie2021quasi}, dose response curves \cite{singh2020kernel}, and time-varying dose response curves \cite{singh2021sequential}. Another strand of this literature incorporates instrumental variables to deal with unobserved confounding \cite{carrasco2007linear,singh2019kernel}. We study a different problem that combines short term experimental data with long term observational data. 
 Previous methods using kernels for continuous actions do not handle the complex linkage between short term and long term effects across data sources, and therefore cannot analyze long term causal inference. Our contribution is a method to do so. 
 
Prior to this work, \cite{battocchi2021estimating} allow for continuous actions but impose restrictive linearity and homogeneity conditions, in both of the responses and in their link. By contrast, we study a nonparametric dose response estimator with nonlinearity and heterogeneity in the responses and their link.
Subsequently to this paper's circulation \cite{singh2022generalized}, \cite{zeng2024continuous} provide complementary results for the long term dose response curve. 
For a given action value, the authors prove pointwise asymptotic consistency of a pseudo-outcome estimator, and pointwise asymptotic normality centered at the biased probability limit of a local linear estimator.
By contrast, we prove uniform consistency across action values, with nonasymptotic rates, for a kernel ridge regression estimator.
Uniform inference for kernel ridge regression over general domains is an open question in statistics, so uniform inference for our estimators is an important question for future work. Finally, we note that this paper subsumes a related working paper \cite{singh2021generalized}.

Section~\ref{sec:problem} formalizes long term causal inference as a core challenge for artificial intelligence and program evaluation.
Section~\ref{sec:rkhs} summarizes RKHS techniques and assumptions, which serve as our statistical framework.
Section~\ref{sec:method} uses RKHS techniques to derive a closed form estimator, with interpretable weights for extrapolating long term effects.
Section~\ref{sec:theory} uses RKHS assumptions to derive uniform rates of convergence that reflect the effective dimension of the data.
Section~\ref{sec:application} demonstrates that our long term dose response estimators recover the long term effects of continuous actions reasonably well, using real data.

\section{Problem: Long term causal inference}\label{sec:problem}

\subsection{Goal: Long term dose response curve}

\begin{figure}
\begin{center}
        \includegraphics[width=0.48\textwidth]{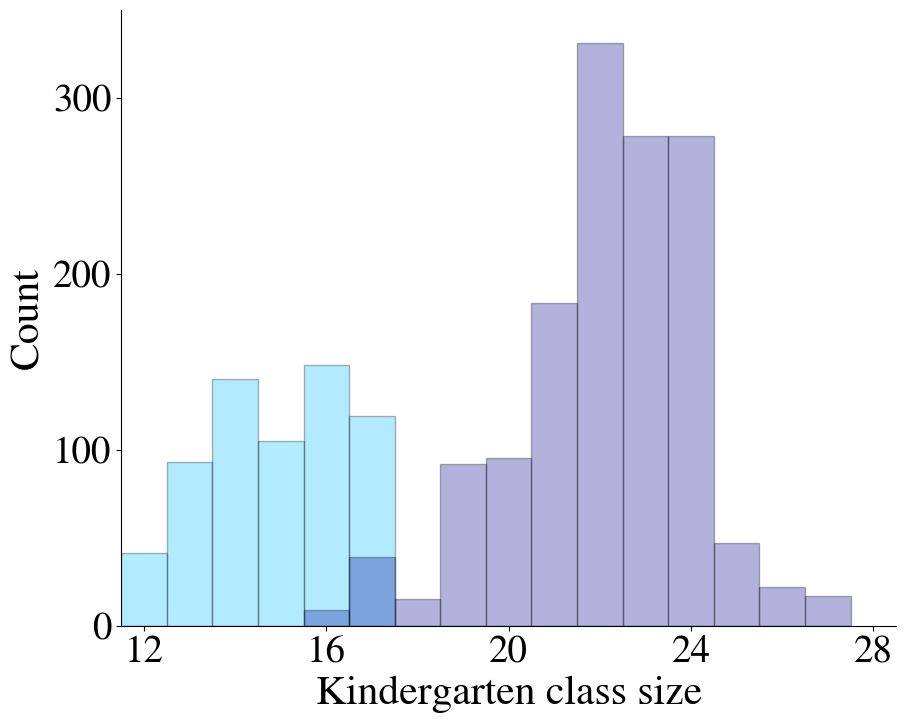}
    \end{center}
    \vspace{-10pt}
    \caption{Real data are reasonably modeled as continuous actions. Much empirical work focuses on ``small'' (light blue) versus ``large'' (dark blue) class sizes, yet the supports are rich and overlapping.}\label{fig:actions} 
\end{figure}

A long term dose response curve is a causal function that summarizes the expected, counterfactual, long term reward $Y^{(d)}$,  given a hypothetical intervention on a continuous action that sets $D=d$. 
For example, Figure~\ref{fig:actions} shows that Project STAR randomly assigned students to many different class sizes, which we model as a continuous action.
The crux of long term causal inference is that we do not have joint observations of the randomized action and long term reward. Nonetheless, we wish to estimate their causal relationship from short term experimental data $(G=\EXP)$ and long term observational data $(G=\OBS)$. We study several variations of long term dose response curves. Unless noted, expectations are with respect to a superpopulation distribution $\P$ with density $\d\P$.

\begin{definition}[Long term dose response curves]\label{def:dose}
   \begin{enumerate}
       \item Dose response: $\theta_0(d)=\E\{Y^{(d)}\}$ is the counterfactual mean long term  reward, of action $D=d$, for the average context. 
       \item Distribution shifted dose response: $\theta_0^{\DS}(d,\tilde{\P})=\E_{\tilde{\P}}\{Y^{(d)}\}$ is the counterfactual mean long term  reward, given action $D=d$, for an alternative context with distribution $\tilde{\P}$. 
       \item Experimental dose response: $\theta_0^{\EXP}(d)=\E\{Y^{(d)}|G=\EXP\}$ is the counterfactual mean long term reward, given action $D=d$, for the average experimental context. 
       \item Observational dose response: $\theta_0^{\OBS}(d)=\E\{Y^{(d)}|G=\OBS\}$ is the counterfactual mean long term reward, given action $D=d$, for the average observational context. 
   \end{enumerate}
\end{definition}

The dose response curve $\theta_0(d)$ is for a representative context, drawn from the superpopulation of contexts which generate the short term experimental data $(G=\EXP)$ and the long term observational data  $(G=\OBS)$. It may depart from the experimental dose response $\theta_0^{\EXP}(d)$ and the observational dose response $\theta_0^{\OBS}(d)$ due to a selection process: $G$, viewed as a random variable, may not be drawn completely at random. By distinguishing among these functions, we emphasize that the experimental data may be limited  in not only duration but also scope; the long term effect of an action in a representative context may be of interest, yet only short term rewards for select contexts may be collected in exploration.

The distribution shifted dose response curve $\theta_0^{\DS}(d,\tilde{\P})$ concerns transfer learning: though our experimental and observational data are drawn from a superpopulation distribution of contexts $\P$, what would be the dose response for an alternative distribution of contexts $\tilde{\P}$ \cite{pearl2014external}? We will consider a rich form of distribution shift formalized below.

Definition~\ref{def:dose} generalizes from long term dose response curves to long term counterfactual distributions by replacing the $\E$ symbols with $\P$ symbols. Long term counterfactual distributions capture aspects of the long term reward distribution beyond its mean, e.g. its variance and skewness. Appendix~\ref{sec:distribution} extends our results accordingly. This work appears to provide the first nonlinear, nonparametric estimators for long term dose response curves and counterfactual distributions.

\subsection{Previous work: Long term identification}

Several previous works identify long term dose response curves, under a variety of identifying assumptions. For clarity of exposition, we focus on the surrogate model for analysis, then demonstrate how our results apply to related models in Remark~\ref{remark:additional}.
We implement kernel estimators for multiple long term causal inference models in Section~\ref{sec:application}. 

The surrogate model posits the following data generating process. A single observation begins with a context $X$ drawn from the distribution $\P$. Conditional on the context, the selection $G\in\{\EXP,\OBS\}$ is as good as random: $\P(G=\EXP|X)$ is a function of the context $X$, but the actual realization of $G$ is a biased coin toss.  

For experimental data, i.e. if $G=\EXP$, then the action $D$ is also random conditional upon $X$:  the density $\d\P(D|G=\EXP,X)$ is a function of the context $X$, but the actual realization of $D$ is a random draw from the density. Finally, we observe the short term reward $S$ drawn from the density $\d\P(S|D=d,G=\EXP,X)$. From experimental data alone, the counterfactual distribution of context-specific short term effects $\P\{S^{(d)}|G=\EXP,X\}$ is identified as $\P\{S|D=d,G=\EXP,X\}$, by standard arguments, under a further overlap condition for actions across experimental contexts.

To identify long term effects, long term data are necessary. However, these long term data are observational: if $G=\OBS$, then the action $D$ is not randomized. In the simplest version of the model, the action is not observed at all when $G=\OBS$. What we do observe is the short term reward $S$ and long term reward $Y$, so $\P(Y|S,G=\OBS,X)$ is identified.

To combine short and long term data, \cite{prentice1989surrogate,athey2020estimating}, and others, articulate nonparametric surrogacy and comparability assumptions. Surrogacy imposes that the action only affects the long term reward via the short term reward in the experiment: $\P(Y|S,D,G=\EXP,X)=\P(Y|S,G=\EXP,X)$.  Comparability imposes that the long term reward follows a common conditional distribution across the experimental and observational data: $\P(Y|S,G=\OBS,X)=\P(Y|S,G=\EXP,X)$. With overlap, these conditions identify the counterfactual distribution of context-specific long term effects $\P\{Y^{(d)}|X\}$ as $\int \P(Y|s,G=\OBS,X)\d\P(s|d,G=\EXP,X)$. 

From this identifying formula, we determine what kinds of distribution shift can be tolerated. It helps if the alternative distribution of contexts $\tilde{\P}$ preserves $\P(Y|S,G=\OBS,X)$ and $\P(S|D,G=\EXP,X)$. For distribution shift, we impose $\tilde{\P}(Y,S,G=\OBS,X)=\P(Y|S,G=\OBS,X)\tilde{\P}(S,G=\OBS,X)$ and $\tilde{\P}(S,D,G=\EXP,X)=\P(S|D,G=\EXP,X)\tilde{\P}(D,G=\EXP,X)$, again with overlap. Importantly, we tolerate shifts in the observational distribution of short term rewards i.e. $\tilde{\P}(S|G=\OBS,X)\neq \P(S|G=\OBS,X)$, and shifts in the experimental distribution of randomized actions i.e. $\tilde{\P}(D|G=\EXP,X)\neq \P(D|G=\EXP,X)$. This extension, to long term transfer learning, appears to be a modest innovation.

\begin{lemma}[Surrogate formula; e.g. Theorem 1 of \cite{athey2020estimating}]\label{lemma:identification}
Suppose standard surrogacy conditions hold, as well as distribution shift conditions for the surrogate model, formalized in Appendix~\ref{sec:identification}. Define the regression function $\gamma_0(S,G,X)=\E(Y|S,G,X)$. Then
    \begin{enumerate}
        \item $\theta_0(d)=\int \int \gamma_0(s,G=\OBS,x)\d\P(s|d,G=\EXP,x)\d\P(x)$;
        \item $\theta^{\DS}_0(d)=\int \int\gamma_0(s,G=\OBS,x)\d\P(s|d,G=\EXP,x)\d\tilde{\P}(x)$;
        \item $\theta^{\EXP}_0(d)=\int \int\gamma_0(s,G=\OBS,x)\d\P(s|d,G=\EXP,x)\d\P(x|G=\EXP)$;
        \item $\theta^{\OBS}_0(d)=\int \int\gamma_0(s,G=\OBS,x)\d\P(s|d,G=\EXP,x)\d\P(x|G=\OBS)$.
    \end{enumerate}
\end{lemma}

Lemma~\ref{lemma:identification} expresses each causal function as a double integral of the regression function $\gamma_0$ learned from observational data, integrated according to a conditional distribution learned from experimental data, and then further integrated according to a possibly conditional distribution of contexts. Since $x$ appears in $\gamma_0(s,G=\OBS,x)$, $\P(s|d,G=\EXP,x)$, and $\P(x)$, the components in this integral are coupled, so the long term dose response seems challenging to estimate.

\begin{remark}[The surrogate formula is generally unbounded when actions are continuous]\label{remark:unbounded}
  Let $\L^2$ be the space of square integrable functions.  Define the functional $F:\L^2\mapsto \R$ as $F:\gamma\mapsto \int\int \gamma(s,G=\OBS,x)\d\P(s|d,G=\EXP,x)\d\P(x)$. When the action is continuous, $F$ is generally unbounded, i.e. there does not exists some constant $C<\infty$ such that $F(\gamma)\leq C\|\gamma\|_{\L^2}$ for all $\gamma\in \L^2$ \cite{newey1994asymptotic,van1991differentiable}. The technical challenge introduced by continuous actions is well known is causal inference \cite{van2018cv}.
\end{remark}

\begin{remark}[Alternative long term models]\label{remark:additional}
    Under alternative identifying assumptions, the integrals are similar. For example, the alternative assumptions of missingness-at-random \cite{kallus2020role} or external validity and latent unconfoundedness \cite{athey2020combining}, lead to $\theta_0(d)=\int \int \gamma_0(s,d,G=\OBS,x)\d\P(s|d,G=\EXP,x)\d\P(x)$ where now $\gamma_0(S,D,G,X)=\E(Y|S,D,G,X)$. Our estimation results apply to all of these variations by slightly modifying $\gamma_0$.
\end{remark}

\section{Statistical framework: RKHS}\label{sec:rkhs}

\subsection{Kernel techniques for algorithm derivation}

We use techniques from the RKHS framework to overcome the practical challenge of estimating the double integral in Lemma~\ref{lemma:identification}, and the technical challenge of estimating an unbounded formula in Remark~\ref{remark:unbounded}. These algorithmic techniques lead to a simple estimator with a closed form solution.

A scalar-valued RKHS $H$ is a Hilbert space of functions $f:\cA\rightarrow\R$, where $\cA$ is some Polish space. Functions in this space are characterized by the feature map, which takes a point $a\in\cA$ and maps it to  features $\phi(a)$ in the RKHS. The closure of $\Span\{\phi(a)\}_{a\in\cA}$ is $H$; the features are the dictionary of basis functions for functions in $H$. In particular, to evaluate $f\in H$ at $a$, we take the RKHS inner product $f(a)=\langle f,\phi(a)\rangle_H$. The kernel $k:\cA\times \cA \rightarrow \R$ is the inner product of features: $k(a,a')=\langle \phi(a),\phi(a')\rangle_H $. It is continuous, symmetric, and positive definite.

The RKHS is widely used for nonparametric regression of some output $Y\in\R$ on some input $a\in\cA$. A kernel ridge regression estimator of the regression function $f_0(A)=\E(Y|A)$ is
\begin{equation}\label{eq:loss}
   \hat{f}=\argmin_{f\in H} \frac{1}{n}\sum_{i=1}^n \{Y_i-\langle f,\phi(A_i) \rangle_H\}^2+\lambda\|f\|_H^2,
\end{equation}
where $\lambda>0$ is a ridge penalty hyperparameter. The minimizer has a well known closed form solution \cite{kimeldorf1971some}, as a weighting $\alpha\in \R^n$ of the output vector $Y\in \R^n$:
\begin{equation}\label{eq:closed}
    \hat{f}(a)=\frac{1}{n}\sum_{i=1}^n Y_i\alpha_i=\frac{1}{n}Y^{\top}\alpha,\quad \alpha=n(K_{AA}+n\lambda I)^{-1}K_{Aa}.
\end{equation}
Here, $K_{AA}\in \R^{n\times n}$ is the kernel matrix with $(i,j)$th entry $k(A_i,A_j)$, and $K_{Aa}\in\R^n$ is the kernel vector with $i$th entry $k(A_i,a)$. 
For tuning $\lambda$, previous work has shown that generalized cross validation and leave-one-out cross validation have closed form solutions, the former is asymptotically optimal, and they are similar in practice \cite{craven1978smoothing,li1986asymptotic}.
We will derive interpretable weights for long term rewards, calculated from the short term experiment, which deliver long term dose response curves.

To take the expectation of a function $f\in H$, a useful technique is the kernel mean embedding:
$$
\E\{f(A)|B=b\}=\int \langle f,\phi(A) \rangle_H\d\P(a|b)=\left\langle f,\int \phi(a)\d\P(a|b) \right\rangle = \langle f, \mu_a(b) \rangle_H,  
$$
where $\mu_a(b)=\int \phi(a)\d\P(a|b)$ embeds the conditional distribution $\P(a|b)$ using the features $\phi(a)$ \cite{smola2007hilbert}. Exchanging the expectation and inner product uses weak regularity conditions (Assumption~\ref{assumption:RKHS} below). In summary, the expectation of a function in the RKHS equals the product of the function and the mean embedding.
We will embed the short term experimental data distribution appropriately, for the extrapolation of long term effects.

The mean embedding $\mu_a(b)=\E \{\phi(A)|B=b\}$ is a generalized regression function, so we will estimate it in a manner that generalizes~\eqref{eq:loss} and~\ref{eq:closed}. In particular, we will estimate its equivalent conditional expectation operator $E_0:f(\cdot)\mapsto \E\{f(A)|B=(\cdot)\}$, which we assume maps from an RKHS $H_{\cA}$ with features $\phi(a)$ to another RKHS $H_{\cB}$ with features $\phi(b)$, for simplicity.\footnote{Formally, $H_{\cA}$ has the kernel $k_{\cA}(a,a')$ and features $\phi_{\cA}(a)$, while $H_{\cB}$ has the possibly different kernel $k_{\cB}(b,b')$ and features $\phi_{\cB}(b)$. We suppress the subscripts to lighten notation.} This ensures that the expectation of a function of $A$, conditional upon $B$, is a smooth function of $B$. It leads to the simple equivalence $\mu_a(b)=E_0^*\phi(b)$, where $E_0^*$ is the adjoint of $E_0$. We estimate $\hat{E}$ as a generalized kernel ridge regression in the space of Hilbert-Schmidt operators $\cL_2(H_{\cA},H_{\cB})$ \cite{micchelli2005learning}, and hence $\hat{\mu}_a(b)=\hat{E}^*\phi(b)$. 
This generalized regression is how we will learn the appropriate representation of the short term experiment for long term causal inference.

For functions defined over multiple variables, we use product kernels. Consider the function $f:\cA\times\cB\rightarrow\R$ defined over $A\in\cA$ and $B\in\cB$. We define an RKHS $H$ for this function by combining RKHSs $H_{\cA}$ and $H_{\cB}$ containing functions $f_{\cA}:\cA\rightarrow \R$ and  $f_{\cB}:\cB\rightarrow \R$, respectively. Its kernel is the product of the individual kernels: $k(a,b;a',b')=k_{\cA}(a,a')k_{\cA}(b,b')$. If $k_{\cA}$ and $k_{\cB}$ are Gaussian kernels, then so is their product, making it a natural choice. Because we use product kernels, the estimators will have the symbol $\odot$ for the elementwise product of kernel matrices. Formally, $H=H_{\cA}\otimes H_{\cB}$ is a tensor product RKHS with features $\phi(a,b)=\phi(a)\otimes \phi(b)$, where the symbols $\otimes$ means outer product: $(x\otimes y)z=x\langle y,z\rangle$. For any $f\in H$, $f(a,b)=\langle f,\phi(a)\otimes \phi(b) \rangle_H$ and $\|\phi(a,b)\|_H=\|\phi(a)\|_{H_{\cA}}\|\phi(b)\|_{H_{\cB}}$. \cite[Section 4]{berlinet2011reproducing} give background.

\subsection{Main assumptions: Low effective dimension and smoothness}\label{sec:assumptions}

Our main assumption is that the RKHS used in estimation has a low effective dimension: while the features may be infinite dimensional, relatively few feature dimensions explain most of the variation in the data. We also assume that the estimands are smooth with respect to these features. In what follows, we formalize these familiar assumptions from RKHS analysis \cite{caponnetto2007optimal,fischer2017sobolev} in terms of the spectrum of the kernel.

To begin, we express the features as a possibly infinite expansion of the kernel. Let $\L^2$ denote the space of square integrable functions from $\cA$ to $\R$ with respect to measure $\P$. For a kernel $k$, define the convolution operator $L:\L^2\rightarrow\L^2$ as $L:f\mapsto \int k(a,\cdot)f(a)\d\P(a)$, which may be expressed in terms of its spectrum: $Lf=\sum_{j=1}^{\infty}\eta_j\langle f,\varphi_j\rangle_H\varphi_j$. Here, $(\eta_j)$ are weakly decreasing eigenvalues, and $(\varphi_j)$ are corresponding eigenfunctions, of the data passed through the kernel. For simplicity, suppose $(\varphi_j)$ form an orthonormal basis of $\L^2$. With Gaussian data and the Gaussian kernel, $(\varphi_j)$ are weighted Hermite polynomials \cite[Section 4.3]{williams2006gaussian}. In general,
$$
\L^2=\left(f=\sum_{j=1}^{\infty}f_j\varphi_j:\sum_{j=1}^{\infty}f_j^2<\infty\right),\quad H=\left(f=\sum_{j=1}^{\infty}f_j\varphi_j:\sum_{j=1}^{\infty}\eta_j^{-1}f_j^2<\infty\right).
$$
The RKHS $H$ is the subset of $\L^2$ for which higher order terms in the series $(\varphi_j)$ have a smaller contribution, as penalized by the eigenvalues $(\eta_j)$. By Mercer's theorem, the feature map is the infinite dictionary of basis functions given by the spectrum of the kernel: $\phi(a)=\{\eta_j^{1/2}\varphi_j(a)\}_{j=1}^{\infty}$. 

\begin{figure}
\captionsetup[subfigure]{justification=Centering}
\begin{subfigure}[t]{0.48\textwidth}
         \centering
        \resizebox{\textwidth}{!}{%
       \includegraphics[width=\textwidth]{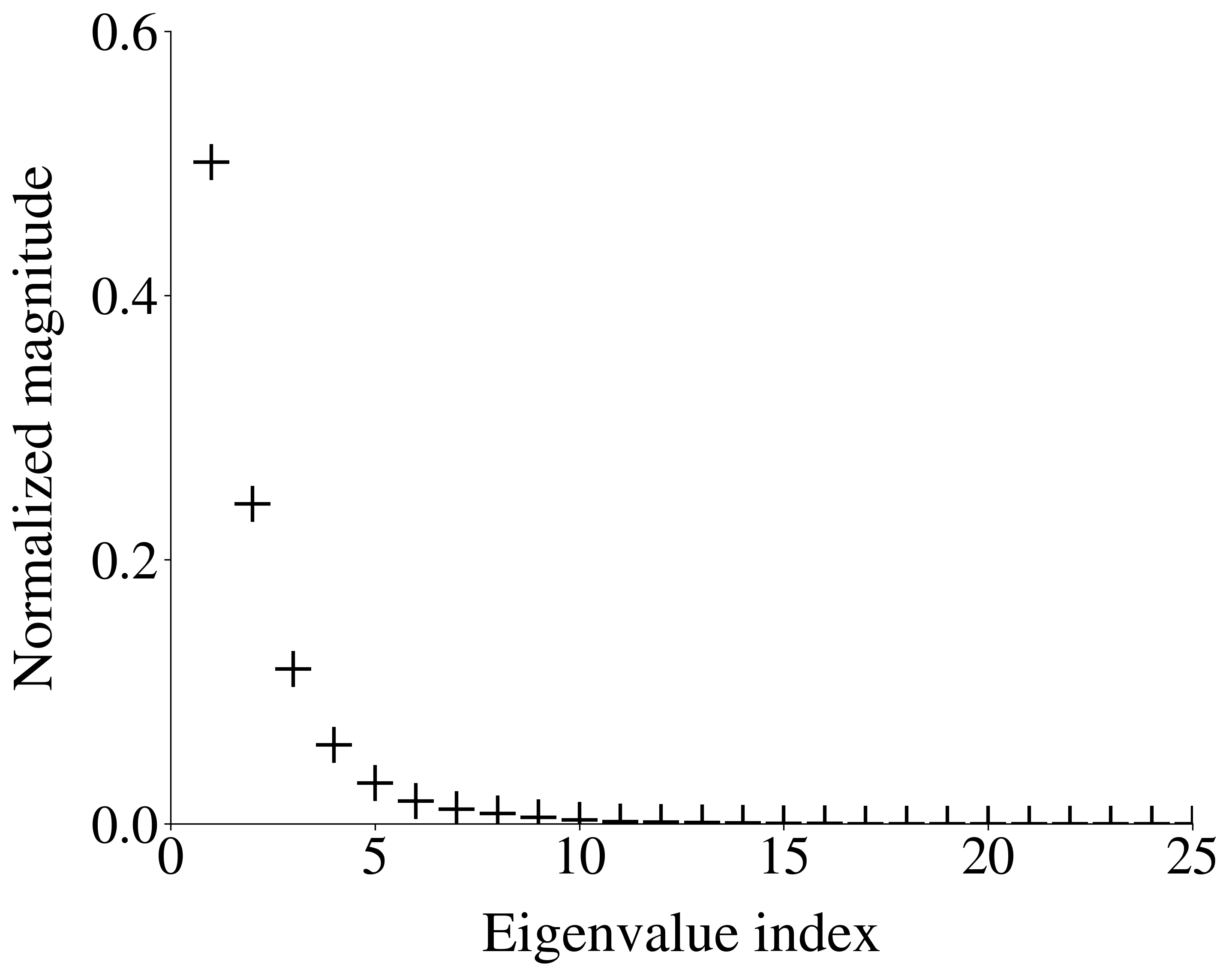}
        }
    \caption{ Assumption~\ref{assumption:smooth}: $b^{\OBS}\geq 1$.}\label{fig:eigen1}
\end{subfigure}\hspace{\fill} 
\begin{subfigure}[t]{0.48\textwidth}
          \centering
        \resizebox{\textwidth}{!}{%
      \includegraphics[width=\textwidth]{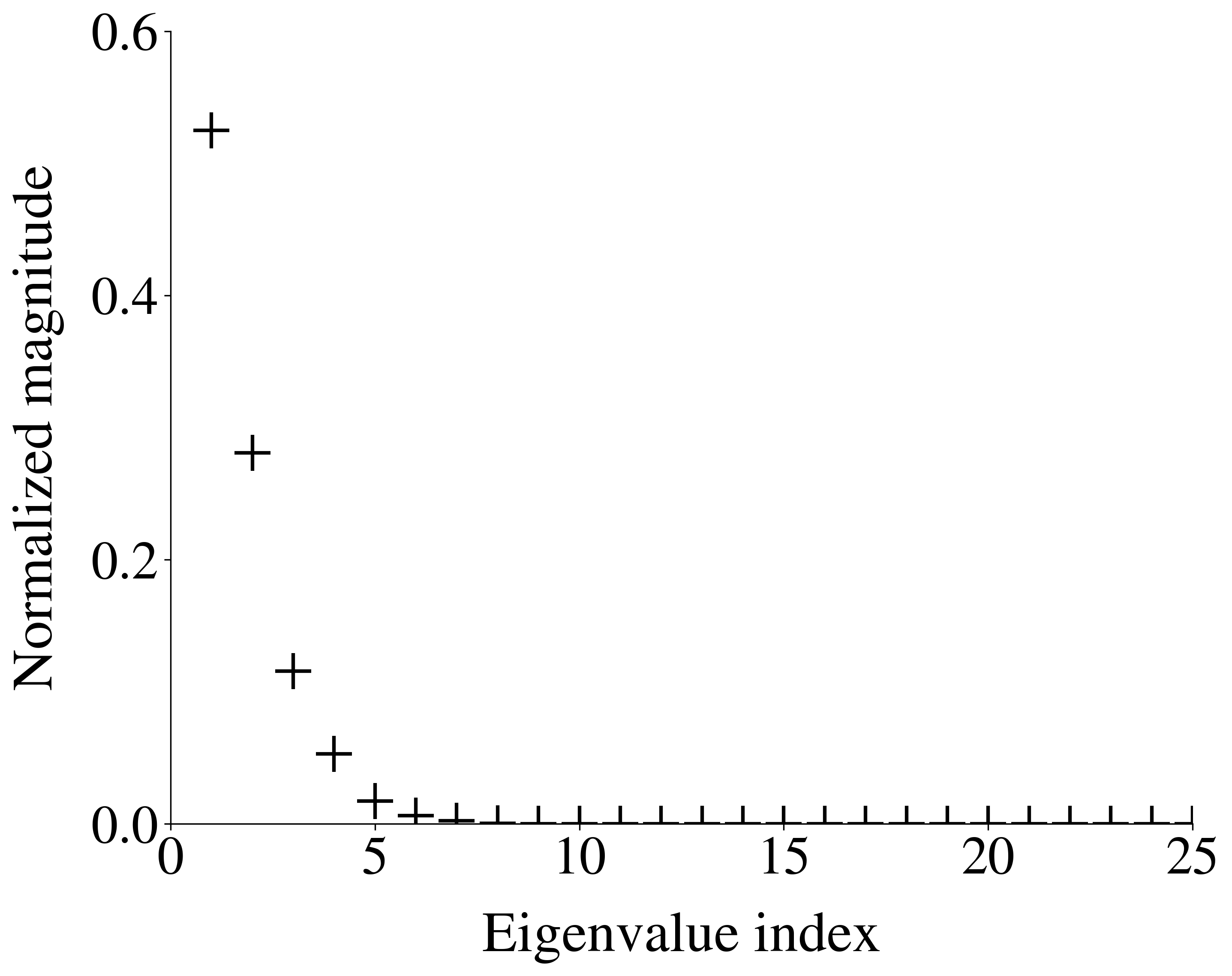}
        }
    \caption{ Assumption~\ref{assumption:smooth_op}: $b^{\EXP}\geq 1$.}\label{fig:eigen2}
\end{subfigure}
\caption{Real data have a low effective dimension in the sense of~\eqref{eq:b}. Figures~\ref{fig:eigen1} and~\ref{fig:eigen2} visualize the initial $25$ eigenvalues of $K_{S^{\OBS}S^{\OBS}}\odot K_{X^{\OBS}X^{\OBS}}$ and $K_{D^{\EXP}D^{\EXP}}\odot K_{X^{\EXP}X^{\EXP}}$, defined in Algorithm~\ref{algo:dose}, using Project STAR data. These are empirical analogues of Assumptions~\ref{assumption:smooth} and~\ref{assumption:smooth_op}.}\label{fig:eigen}
\end{figure}

Our main approximation assumption is that the features have a low effective dimension:
\begin{equation}\label{eq:b}
    \eta_j\leq Cj^{-b},\quad C<\infty,\quad b\geq 1.
\end{equation}
The eigenvalues decay at least polynomially. Any bounded kernel satisfies~\eqref{eq:b} \cite[Lemma 10]{fischer2017sobolev}. A higher value of $b$ corresponds to a lower effective dimension, better control of the variance of our estimator, and hence a faster rate. The limit $b\rightarrow \infty$ gives an RKHS with finite dimension \cite{caponnetto2007optimal}. The empirical eigenvalues are simple to compute, so it is simple to validate this assumption with a diagnostic plot. Figure~\ref{fig:eigen} verifies polynomial decay of the empirical eigenvalues in the real world application of Section~\ref{sec:application}; the Project STAR data have a low effective dimension as required by Assumptions~\ref{assumption:smooth} and~\ref{assumption:smooth_op}.\footnote{Specifically, we divide each empirical eigenvalue by the trace of the corresponding matrix, to convey the fraction of variation explained.}

\begin{figure}
\centering 
\captionsetup[subfigure]{justification=Centering}
\begin{subfigure}[t]{0.48\textwidth}
         \centering
        \resizebox{\textwidth}{!}{%
       \includegraphics[width=\textwidth]{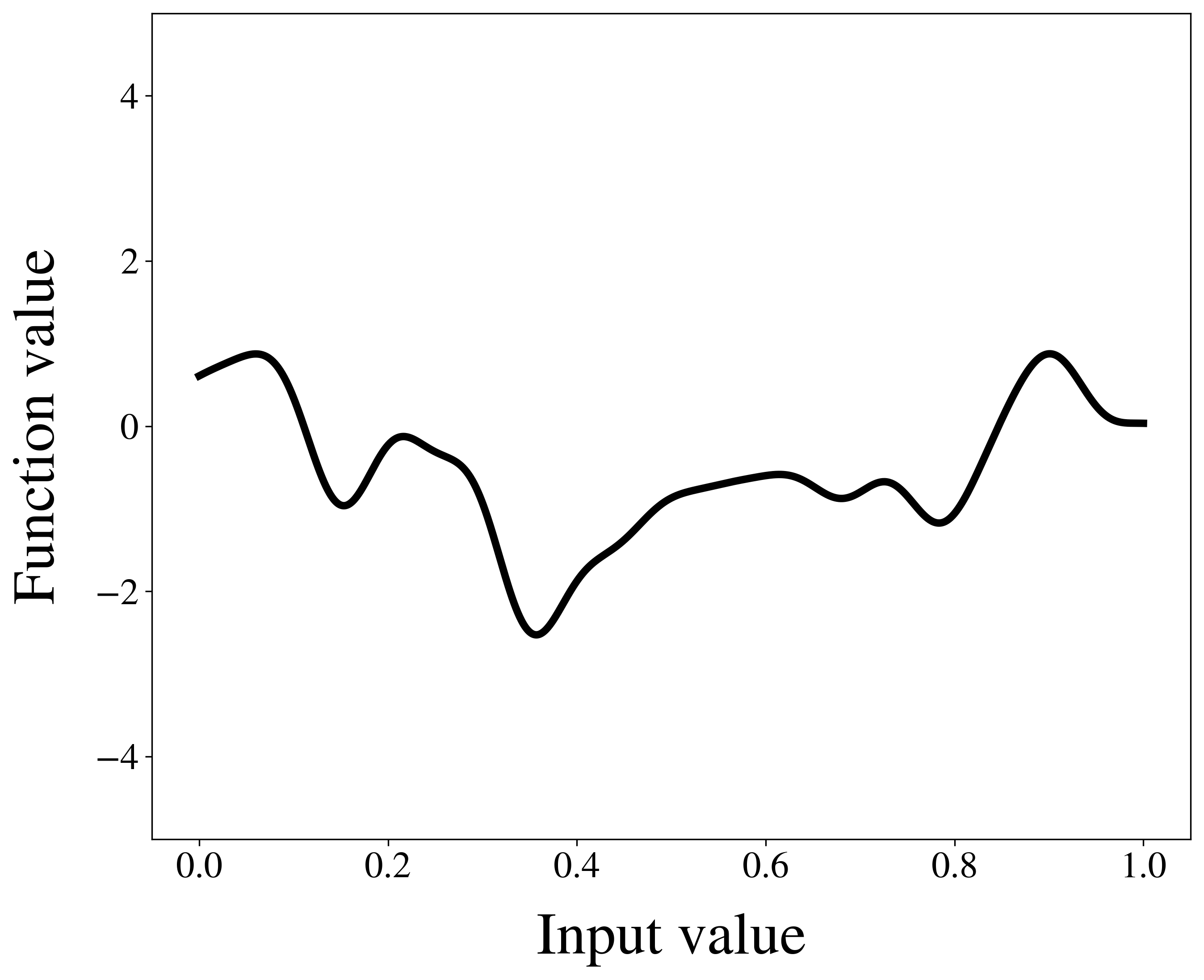}
        }
    \caption{Source condition with $c=1$.}\label{fig:source1}
\end{subfigure}\hfill  
\begin{subfigure}[t]{0.48\textwidth}
          \centering
        \resizebox{\textwidth}{!}{%
      \includegraphics[width=\textwidth]{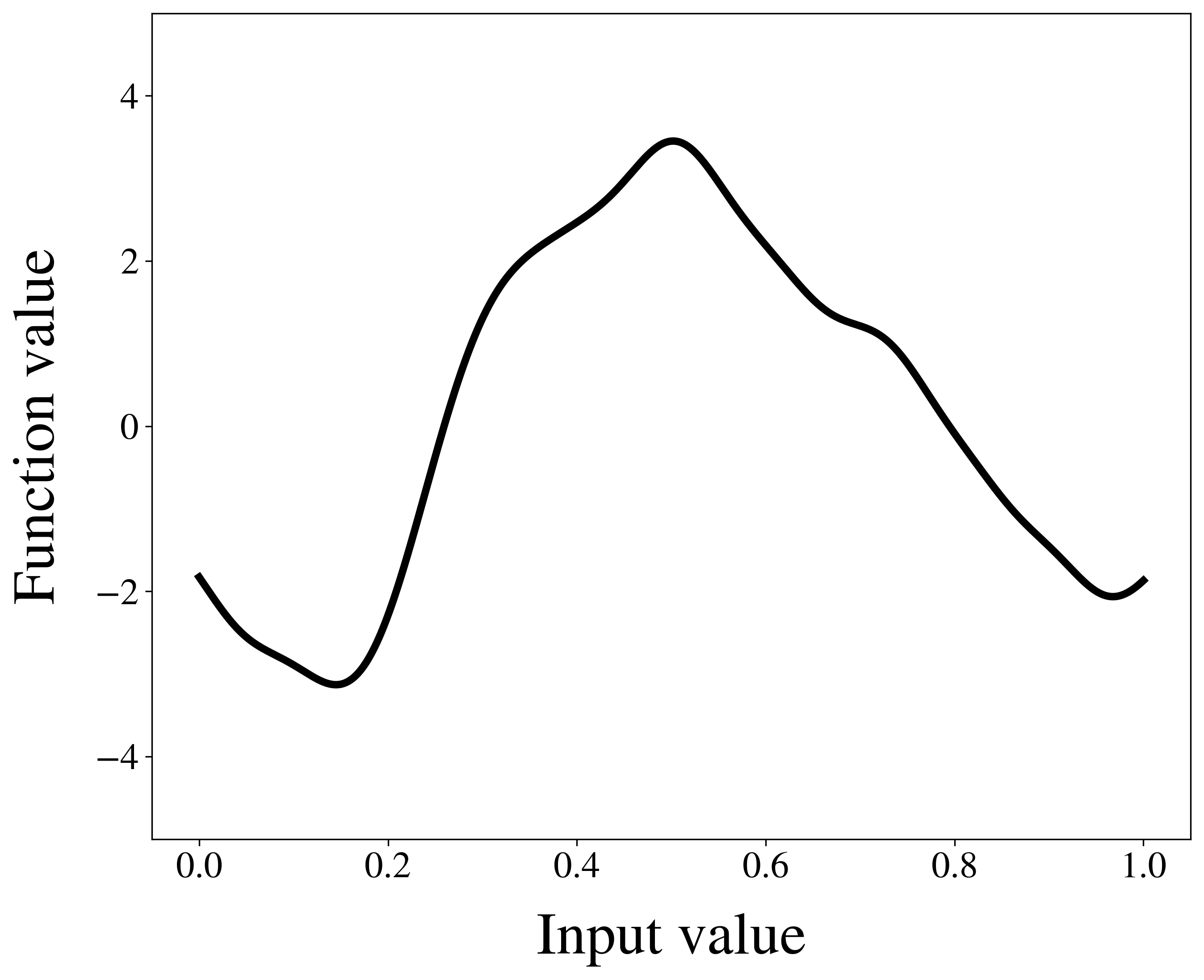}
        }
    \caption{Source condition with $c=2$.}\label{fig:source2}
\end{subfigure}
\caption{The source condition in Sobolev space means the number of square integrable derivatives. For illustration, we take $H$ to be the Sobolev space over $\mathbb{R}$ with one square integrable derivative, with respect to Lebesgue measure, so $b=2$ in the sense of~\eqref{eq:b}. Figures~\ref{fig:source1} and~\ref{fig:source2} visualize $f_0\in H^1$ versus $f_0\in H^2$, i.e. $c=1$ versus $c=2$ in the sense of~\eqref{eq:c}.}\label{fig:sources}
\end{figure}

We also assume that the estimands are smooth respect to these features, satisfying what is known as a source condition \cite{smale2005shannon,carrasco2007linear}:
\begin{equation}\label{eq:c}
    f_0\in H^c=\left(f=\sum_{j=1}^{\infty}f_j\varphi_j:\sum_{j=1}^{\infty}\eta_j^{-c}f_j^2<\infty\right),\quad c\in(1,2].
\end{equation}
Note that $H^0=\L^2$ and $H^1=H$, while $H^c$ is contained in $H$ for $c>1$. A larger value of $c$ corresponds to a smoother target $f_0$, better control of the bias of our estimator, and hence a faster rate. Figure~\ref{fig:sources} visualizes $c=1$ versus $c=2$ in a concrete example.

These approximation assumptions generalize well known Sobolev conditions \cite{fischer2017sobolev}. Consider the Sobolev space  $\H_2^s$ of functions $f:\cA\rightarrow \R$ with $\cA\subset\R^p$, where $s$ is the number of square integrable derivatives, with respect to the Lebesgue measure. This space is an RKHS when $s>p/2$, with the Mat\'ern kernel \cite[Theorem 143]{berlinet2011reproducing}. Take $H=\H_2^s$ with $s>p/2$ as the RKHS for estimation. If the target $f_0$ has $s_0$ derivatives, i.e. $f_0\in \H_2^{s_0}$, then $c=s_0/s$ and hence $\H_2^{s_0}=(\H_2^s)^c$ in the sense of~\eqref{eq:c}. Meanwhile, the effective dimension of  $\H_2^s$ is $b=2s/p$ in the sense of~\eqref{eq:b}. Intuitively, $c$ measures the relative smoothness of the target to the estimator, while $b$ measures the relative smoothness of the estimator to the dimension. As $s\rightarrow \infty$, the Mat\'ern kernel converges to the widely used Gaussian kernel.

Rates in terms of $(b,c)$ reflect the effective dimension of the features, and the smoothness of the estimand relative to the features. The minimax optimal rate for regression in RKHS norm is $n^{-\frac{1}{2}\frac{c-1}{c+1/b}}$, achieved by kernel ridge regression with rate optimal regularization $\lambda=n^{-\frac{1}{c+1/b}}$. This rate approaches $n^{-1/4}$ when $b\rightarrow \infty$ and $c=2$, i.e. when the effective dimension is finite and the smoothness is high. This paper's results combine rates (and regularization paths) of this form. They apply to Sobolev spaces over $\R^p$ as a special case, where $n^{-\frac{1}{2}\frac{c-1}{c+1/b}}=n^{-\frac{s_0-s}{2s_0+p}}$. More generally, they allow variables to belong to Polish spaces.
\section{Method: Long term dose response via kernels}\label{sec:method}

\subsection{Short term kernel embedding}

We construct an RKHS for long term dose response curve estimation, in which we embed the distribution of the short term experimental data. We define RKHSs for the short term reward $S \in\cS$, action $D\in\cD$, and context $X \in\cX$.
For example, for the short term reward, we define the RKHS $H_{\cS}$ with feature map $\phi_{\cS}(s)$ and kernel $k_{\cS}(s,s')=\langle \phi_{\cS}(s),\phi_{\cS}(s')\rangle_{H_{\cS}}$.

\begin{remark}[Subscripts]
    To lighten notation, we suppress subscripts when arguments are provided, e.g. we write $\phi(s)=\phi_{\cS}(s)$ and $\phi(x)=\phi_{\cX}(x)$, though the feature maps may differ.
\end{remark}

From these RKHSs for short term rewards and contexts, we construct an RKHS $H$ for the long term reward mechanism $\gamma_0^{\OBS}(S,X)=\gamma_0(S,G=\OBS,X)=\E(Y|S,G=\OBS,X)$. We assume that it belongs to the RKHS with the product kernel $k(s,x;s',x')=k_{\cS}(s,s')k_{\cX}(x,x')$. Equivalently, $\gamma_0^{\OBS}\in H=H_{\cS} \otimes H_{\cX}$, where $\otimes$ means tensor product.\footnote{For the alternative long term causal inference models in Remark~\ref{remark:additional}, $H=H_{\cS} \otimes H_{\cD}\otimes H_{\cX}$ instead.} The features for $\gamma_0^{\OBS}$ are the outer product of features for the individual RKHSs: $\gamma_0^{\OBS}(s,x)=\langle \gamma_0^{\OBS},\phi(s,x) \rangle_{H}=\langle \gamma_0^{\OBS},\phi(s)\otimes \phi(x) \rangle_{H}$.

\begin{example}[Quadratic kernels]\label{ex:quadratic}
    As a toy example, consider $\cS,\cX\subset \R$ and quadratic kernels, e.g. $k_{\cS}(s,s')=(ss'+1)^2$. Each kernel has three features, e.g. $\phi(s)=(s^2,\sqrt{2}s,1)$. Hence the product kernel has $3\times 3$ induced features; $\phi(s,x)$ contains each interaction among $(s^2,\sqrt{2}s,1)$ and $(x^2,\sqrt{2}x,1)$.
\end{example}

\begin{example}[Gaussian kernels]\label{ex:gaussian}
    With Gaussian data and Gaussian kernels, the features of each kernel are weighted Hermite polynomials, of which there are countably many. In this richer example, $\phi(s,x)$ interacts the Hermite polynomials of $s$ and $x$. This basis allows for nonlinearity and heterogeneity in the link between short and long term effects. It satisfies our regularity conditions.
\end{example}

\begin{assumption}[RKHS regularity]\label{assumption:RKHS}
    \begin{enumerate}
    \item The features $\phi(S)$, $\phi(D)$, and $\phi(X)$ are measurable.
        \item The kernels $k_{\cS}$, $k_{\cD}$, and $k_{\cX}$ are bounded: $\sup_{s\in\cS}\|\phi(s)\|_{H_{\cS}}\leq\kappa_s$, $\sup_{d\in\cD}\|\phi(d)\|_{H_{\cD}}\leq\kappa_d$, and $\sup_{x\in\cX}\|\phi(x)\|_{H_{\cX}}\leq\kappa_x$.
        \item The kernels $k_{\cS}$ and $k_{\cX}$  are characteristic: for all Borel probability measures $\Q(s)$ and $\Q(x)$, the mappings $\Q(s)\mapsto \int \phi(s)\d\Q(s)$ and $\Q(x)\mapsto \int \phi(x)\d\Q(x)$ are injective.
    \end{enumerate}
\end{assumption}

Measurability, boundedness, and the characteristic property are weak regularity conditions satisfied by commonly used kernels, e.g. the Mat\'ern and Gaussian kernels. The characteristic property ensures that kernel mean embeddings of distributions are unique \cite{sriperumbudur2010relation}.

\begin{theorem}[Representation via short term kernel embedding]\label{theorem:representation}
    If Assumption~\ref{assumption:RKHS} holds, then the short term kernel embedding $\mu^{\EXP}_s(d,x)=\int \phi(s)\d\P(s|d,G=\EXP,x)$ exists. If, in addition, the conditions of Lemma~\ref{lemma:identification} hold and $\gamma_0^{\OBS}\in H$, then we have the representations
    \begin{enumerate}
        \item $\theta_0(d)=\langle \gamma_0^{\OBS}, \mu_{s,x}(d) \rangle_H$ where $\mu_{s,x}(d)=\int \{\mu_s^{\EXP}(d,x) \otimes \phi(x)\}\d\P(x)$;
        \item  $\theta^{\DS}_0(d)=\langle \gamma_0^{\OBS}, \tilde{\mu}_{s,x}(d) \rangle_H$ where $\tilde{\mu}_{s,x}(d)=\int \{\mu_s^{\EXP}(d,x) \otimes \phi(x)\}\d\tilde{\P}(x)$;
        \item $\theta^{\EXP}_0(d)=\langle \gamma_0^{\OBS}, \mu'_{s,x}(d) \rangle_H$ where $\mu'_{s,x}(d)=\int \{\mu_s^{\EXP}(d,x) \otimes \phi(x)\}\d\P(x|G=\EXP)$;
        \item $\theta^{\OBS}_0(d)=\langle \gamma_0^{\OBS}, \mu''_{s,x}(d) \rangle_H$ where $\mu''_{s,x}(d)=\int \{\mu_s^{\EXP}(d,x) \otimes \phi(x)\}\d\P(x|G=\OBS)$.
    \end{enumerate}
\end{theorem}

\begin{proof}[Proof sketch]
By Lemma~\ref{lemma:identification}, $\theta_0(d)=\int \int \gamma_0^{\OBS}(s,x)\d\P(s|d,G=\EXP,x)\d\P(x)$, which equals
$$
\int \int \langle \gamma^{\OBS}_0,\phi(s,x)\rangle_H \d\P(s|d,G=\EXP,x)\d\P(x)=\left\langle \gamma^{\OBS}_0,\int \int \phi(s,x)\d\P(s|d,G=\EXP,x)\d\P(x)\right\rangle_H.
$$
Using our tensor product construction and Assumption~\ref{assumption:RKHS}, we express the double integral as
$$
\int \int \{\phi(s) \otimes \phi(x)\} \d\P(s|d,G=\EXP,x)\d\P(x)=\int  \left[\left\{\int \phi(s)\d\P(s|d,G=\EXP,x)\right\} \otimes \phi(x)\right] \d\P(x). \qedhere 
$$
\end{proof}

The long term regression $\gamma_0^{\OBS}(s,x)$ allows for nonlinearity and heterogeneity in the link 
between the short term and long term. 
The short term kernel mean embedding $\mu^{\EXP}_s(d,x)$ allows for nonlinearity and heterogeneity in the counterfactual distribution of short term rewards.  
This embedding is a representation of the experiment, and we must learn it from the data. The ``fused'' embeddings $\mu_{s,x}(d)$, $\tilde{\mu}_{s,x}(d)$, $\mu'_{s,x}(d)$, and $\mu''_{s,x}(d)$ further adjust for the possibility that experiments are limited in scope, i.e. that the distribution of experimental contexts $\P(X|G=\EXP)$ is not representative.

\begin{remark}[Embeddings that fuse experimental and observational data]
    Here, $\mu''_{s,x}(d)$ fuses representations of the short term experiment $\mu^{\EXP}(d,x)$ and the context $\phi(x)$ according to the observational distribution $\P(x|G=\OBS)$. Fusing experimental and observational data into the same embedding is necessary to adjust for the limited duration and scope of experiments. 
    Our fused representation for long term dose responses departs from representations of other problems.
\end{remark}

For intuition, we interpret Theorem~\ref{theorem:representation} in terms of Example~\ref{ex:quadratic}. Here, $\mu_s^{\EXP}(d,x)$ encodes short term experimental data $\P(S|D,G=\EXP,X)$ as the conditional moments $\int s^2\d\P(s|D,G=\EXP,X)$, $\sqrt{2}\int s \d\P(s|D,G=\EXP,X)$, and $1$. The fused embedding $\mu_{s,x}(d)$ encodes the desired context distribution $\P(X)$ by interacting these conditional moments with the context moments.\footnote{Altogether, there are $3\times 3$ moments: 
$\int x^2\int s^2\d\P(s|d,G=\EXP,x) \d\P(x)$, $\sqrt{2}\int x^2\int s \d\P(s|d,G=\EXP,x) \d\P(x)$, and $\int x^2 \d\P(x)$; 
$\sqrt{2}\int x\int s^2\d\P(s|d,G=\EXP,x) \d\P(x)$, $2\int x\int s \d\P(s|d,G=\EXP,x) \d\P(x)$, and $\sqrt{2}\int x \d\P(x)$;
and $\int\int s^2\d\P(s|d,G=\EXP,x) \d\P(x)$, $\sqrt{2}\int\int s \d\P(s|d,G=\EXP,x) \d\P(x)$, and $1$.} In this toy example, the long term dose response $\theta_0(d)$ is simply the long term regression $\gamma_0^{\OBS}$ evaluated at these interacted quadratic moments. Quadratic kernels only preserve the first and second moments of the data, so we use characteristic kernels to preserve all moments.

\begin{remark}[The surrogate formula is bounded over $H$ when actions are continuous]\label{remark:bounded}
    Define the functional $F:H\rightarrow \R$ as $F:\gamma\mapsto \int\int \gamma(s,x)\d\P(s|d,G=\EXP,x)\d\P(x)$. Under Assumption~\ref{assumption:RKHS}, we prove that $F$ is bounded even when the action is continuous: there exists a constant $C<\infty$ such that $F(\gamma)\leq C\|\gamma\|_H$ for all $\gamma\in H$. 
    This generalizes the insight of \cite{singh2020kernel}, who study dose response curves when experimental actions and final rewards are jointly observed. 
    Specifically, we use the definition of the RKHS as the space of functions for which the functional $\gamma\mapsto \gamma(s,x)$ is bounded \cite{berlinet2011reproducing}. Boundedness of the functional $F$ implies existence of the short term kernel embedding $\mu_s^{\EXP}(d,x)$ and the fused embeddings.
\end{remark}

    The product kernel is not necessary for Theorem~\ref{theorem:representation}. For example, the same result holds for the sum kernel, replacing the tensor product $\otimes$ with the concatenation $\oplus$. However, the product kernel is interpretable as interactions that allow for heterogeneity in the link and the effects. It also simplifies the proof of Theorem~\ref{theorem:consistency} below. Further products, besides $\phi(s,x)=\phi(s)\otimes \phi(x)$, are only for the sake of exposition. Moreover, measurability of $\phi(D)$ and boundedness of $k_{\cD}$ are not necessary for Theorem~\ref{theorem:representation} in the surrogate model, but they are invoked in the alternative long term models of Remark~\ref{remark:additional}.

\subsection{Closed form solution}

Theorem~\ref{theorem:representation} provides a blueprint for a long term dose response curve algorithm that avoids density estimation and numerical integration. Our algorithm is of the form $\hat{\theta}(d)=\langle \hat{\gamma}^{\OBS}, \hat{\mu}_{s,x}(d) \rangle_H$ where $\hat{\gamma}^{\OBS}$ is a kernel ridge regression estimated from observational data, $\hat{\mu}^{\EXP}_s(d,x)$ is a generalized kernel ridge regression estimated from experimental data, and $\hat{\mu}_{s,x}(d)=\frac{1}{n}\sum_{i=1}^n \{\hat{\mu}_s^{\EXP}(d,X_i) \otimes \phi(X_i)\}$ averages over all data. We prove that this algorithm has a closed form solution, generalizing~\eqref{eq:closed}.

\begin{algorithm}[Long term dose response curve estimator]\label{algo:dose}
Let the superscript $\EXP$ or $\OBS$ denote which data are being used. 
Let $n=n^{\OBS}+n^{\EXP}$ be the total number of observations from the distribution $\P$, and let $\tilde{n}$ be the number of observations from the alternative context distribution $\tilde{\P}$.
For example, $K_{S^{\OBS}S^{\OBS}}\in \R^{n^{\OBS}\times n^{\OBS}}$ is the kernel matrix for observational short term rewards, with $(i,j)$th entry $k_{\cS}(S_i,S_j)$, where $i,j\in\OBS$. Meanwhile, $K_{S^{\OBS}S^{\EXP}}\in \R^{n^{\OBS}\times n^{\EXP}}$ is the kernel matrix comparing observational and experimental short term rewards, with $(i,j)$th entry $k_{\cS}(S_i,S_j)$, where $i\in\OBS$ and $j\in \EXP$. Let $\odot$ be the elementwise product. The long term dose response estimator is
\begin{align*}
    \hat{\theta}(d)&=\frac{1}{n}\sum_{i=1}^n
(Y^{\OBS})^{\top}(K_{S^{\OBS}S^{\OBS}}\odot K_{X^{\OBS}X^{\OBS}}+n^{\OBS}\lambda^{\OBS} I^{\OBS})^{-1}\\
&[\{K_{S^{\OBS} S^{\EXP}}(K_{D^{\EXP}D^{\EXP}}\odot K_{X^{\EXP}X^{\EXP}}+n^{\EXP}\lambda^{\EXP}I^{\EXP})^{-1}(K_{D^{\EXP}d}\odot K_{X^{\EXP}X_i})\}\odot K_{X^{\OBS}X_i}],
\end{align*}
where $I^{\OBS}$ 
and $I^{\EXP}$ 
are identity matrices. The other estimators are similar. For $\hat{\theta}^{\DS}(d)$, $\hat{\theta}^{\EXP}(d)$, and $\hat{\theta}^{\OBS}(d)$, replace $\frac{1}{n}\sum_{i=1}^n(\cdot)$ with $\frac{1}{\tilde{n}}\sum_{i=1}^{\tilde{n}}(\cdot)$, $\frac{1}{n^{\EXP}}\sum_{i\in \EXP}(\cdot)$, and $\frac{1}{n^{\OBS}}\sum_{i\in \OBS}(\cdot)$, respectively.
\end{algorithm}

\begin{proof}[Derivation sketch]
    Analogously to~\eqref{eq:loss}, the component ridge regression estimators are
    \begin{align*}
        \hat{\gamma}^{\OBS}
        &=\argmin_{\gamma\in H} \frac{1}{n^{\OBS}} \sum_{i\in \OBS} \{Y_i-\langle \gamma, \phi(S_i)\otimes \phi(X_i)\rangle_H \}^2+\lambda^{\OBS} \|\gamma\|_H^2, 
        \\
        \hat{E}^{\EXP}
        &=\argmin_{E\in \cL_2(H_{\cS},H_{\cD}\otimes H_{\cX})} \frac{1}{n^{\EXP}} \sum_{i\in \EXP}[\phi(S_i)-E^*
        \{\phi(D_i)\otimes \phi(X_i)\} ]^2+\lambda^{\EXP} \|E\|_{\cL_2(H_{\cS},H_{\cD}\otimes H_{\cX})}^2,
    \end{align*}
    where $\hat{\mu}_s^{\EXP}(d,x)=(\hat{E}^{\EXP})^*\{\phi(d)\otimes \phi(x)\}$. Analogously to~\eqref{eq:closed}, the closed forms are
    \begin{align*}
        \hat{\gamma}^{\OBS}(\cdot,x)&=(Y^{\OBS})^{\top}(K_{S^{\OBS}S^{\OBS}}\odot K_{X^{\OBS}X^{\OBS}}+n^{\OBS}\lambda^{\OBS} I^{\OBS})^{-1}(K_{S^{\OBS}(\cdot)}\odot K_{X^{\OBS}x}) \\
        \{\hat{\mu}_s^{\EXP}(d,x)\}(\cdot)&=K_{(\cdot) S^{\EXP}}(K_{D^{\EXP}D^{\EXP}}\odot K_{X^{\EXP}X^{\EXP}}+n^{\EXP}\lambda^{\EXP}I^{\EXP})^{-1}(K_{D^{\EXP}d}\odot K_{X^{\EXP}x}).
    \end{align*}
    Match the empty arguments of the ridge estimators then average over the placeholder $x$.
\end{proof}

Theorem~\ref{theorem:consistency} gives theoretical values of $(\lambda^{\OBS},\lambda^{\EXP})$ to optimally balance bias and variance for the regressions. Appendix~\ref{sec:tuning} gives practical tuning procedures with closed form solutions to empirically balance bias and variance, one of which is asymptotically optimal for regression.  

\subsection{Towards trustworthiness: Long term reward weights}

A mature literature on kernel methods for binary actions, with joint observations of randomized actions and final outcomes, emphasizes the interpretation of outcome weights; see e.g. \cite{bruns2023augmented} for a recent review and many references, going back to the classical work of \cite{speckman1979minimax}. We show that our estimators share some of these virtues despite the additional challenges of our setting: continuous actions and long term extrapolation.

\begin{corollary}[Interpretable weights]\label{cor:weights}
    Our estimators may be viewed as long term reward weights: $\hat{\theta}(d)=\frac{1}{n_{\OBS}}\sum_{i\in\OBS} Y_i\hat{\alpha}_i(d)$ for a weight vector $\hat{\alpha}(d)\in\R^{n^{\OBS}}$ that is a byproduct of Algorithm~\ref{algo:dose}.
\end{corollary}

These weights allow us to evaluate how trustworthy our estimator is in certain respects, e.g. its robustness and transparency.
For example, we may compare different choices of the kernels $k_{\cS}$ and $k_{\cX}$ in the nonlinear model $\gamma_0^{\OBS}$ that links the short term and long term. For each choice, we can examine the magnitudes of the weights to assess which regions of the short term reward and context space receive greater weight. A larger $|\hat{\alpha}_i(d)|$ means that observation $i\in\OBS$ has more influence on the final estimate. For each choice, we can also examine the signs of the weights to assess how much extrapolation is taking place. A negative weight, i.e. $\hat{\alpha}_i(d)<0$, means that our counterfactual prediction of the long term reward does not simply interpolate among $Y_i$ for $i\in\OBS$.

Such weights would not exist in general, for long term dose response curve estimators outside of the RKHS. As shown in Remark~\ref{remark:unbounded}, the surrogate formula with continous actions is unbounded over $\L^2$ in general. As such, it does not have a Riesz representer over $\L^2$. The weights, if they exist, are empirical analogues of the Riesz representer.

Our algorithm keenly relies on the fact that such weights do exist for long term dose response curve estimators in the RKHS. As shown in Remark~\ref{remark:bounded}, the surrogate formula with continuous actions is bounded over $H$. As such, it does have a Riesz representer over $H$, given by the fused embedding $\mu_{s,x}(d)$  in Theorem~\ref{theorem:representation}. The weights $\hat{\alpha}(d)$ are calculated from its empirical analogue.
\section{Theory: Uniform consistency with finite sample rates}\label{sec:theory}

For identification in Section~\ref{sec:problem}, we placed the long term causal inference assumptions. To derive a closed form estimator in Section~\ref{sec:method}, we assumed RKHS regularity (Assumption~\ref{assumption:RKHS}). To prove uniform consistency across actions, we now place three final assumptions: original space regularity (Assumption~\ref{assumption:original}), low effective dimension and smoothness of the long term regression (Assumption~\ref{assumption:smooth}) and likewise for the short term conditional expectation operator (Assumption~\ref{assumption:smooth_op}).

\begin{assumption}[Original space regularity]\label{assumption:original} 
\begin{enumerate}
\item The long term reward $Y\in\cY\subset\R$ is bounded.
\item The spaces $\cS$, $\cD$,  and  $\cX$ are Polish, i.e. separable and completely metrizable.
\end{enumerate}
\end{assumption}

For simplicity, the long term reward is a bounded scalar. Our results in Appendix~\ref{sec:distribution} relax this condition, allowing $\cY$ to be separable Hilbert space. We allow short term rewards, actions, and contexts to be discrete or continuous. This generality means our results can be applied to nonstandard data such as texts, images, and graphs

Next, we place our main approximation assumptions for the long term regression and short term conditional expectation operator. We assume that they have low effective dimensions and are smooth in the sense of~\eqref{eq:b} and~\eqref{eq:c}, which are standard assumptions in the RKHS framework. Extending the notation of Section~\ref{sec:assumptions}, we specify the measure and space of the eigendecomposition.\footnote{For example, $\eta^{\OBS}_j(H)$ is the $j$th eigenvalue of the operator $L^{\OBS}:\gamma(\cdot )\mapsto \int k\{s,x;(\cdot)\}\gamma(s,x)\d\P(s,x|G=\OBS)$, where $k$ is the kernel of the RKHS $H=H_{\cS}\otimes H_{\cX}$.}

\begin{assumption}[Effective dimension and smoothness for long term regression]\label{assumption:smooth}   The eigenvalues decay, and the regression satisfies a source condition: 
$\eta^{\OBS}_j(H)\leq C j^{-(b^{\OBS})}$ and $\gamma_0^{\OBS}\in H^{(c^{\OBS})}$, for some $b^{\OBS}\geq 1$ and $c^{\OBS}\in(1,2]$. Here and below, $C<\infty$ is an absolute constant.
\end{assumption}

\begin{remark}[Interpretation and diagnostic]
    We assume that the long term regression $\gamma_0^{\OBS}(S,X)=\E(Y|S,G=\OBS,X)$ is smooth relative to the eigenfunctions of the product kernel $k_{\cS}(s,s')k_{\cX}(x,x')$, and that this kernel has eigenvalues that decay at least polynomially, with respect to $\P(S,X|G=\OBS)$. Figure~\ref{fig:eigen1} provides evidence of polynomial decay using a product of Gaussian kernels and real Project STAR data ($K_{S^{\OBS}S^{\OBS}}\odot K_{X^{\OBS}X^{\OBS}}$ in Algorithm~\ref{algo:dose}). When using a product of Gaussian kernels with Gaussian data, the eigenfunctions are Hermite polynomials of $s$ and $x$.
\end{remark}

\begin{assumption}[Effective dimension and smoothness for short term conditional expectation operator]\label{assumption:smooth_op}  The eigenvalues decay, and the operator  satisfies a source condition:   $\eta^{\EXP}_j(H_{\cD}\otimes H_{\cX})\leq C j^{-(b^{\EXP})}$ and $E_0^{\EXP}\in \cL_2\{H_{\cS},(H_{\cD}\otimes H_{\cX})^{(c^{\EXP})}\}$ for some $b^{\EXP}\geq 1$ and $c^{\EXP}\in(1,2]$. 
\end{assumption}

\begin{remark}[Interpretation and diagnostic]\label{remark:op}
    We assume that the short term generalized regression $\mu_s^{\EXP}(d,x)=\E\{\phi(S)|D=d,G=\EXP,X=x\}=(E_0^{\EXP})^*\phi(d,x)$ is smooth relative to the eigenfunctions of the product kernel $k_{\cD}(d,d')k_{\cX}(x,x')$, and that this kernel has eigenvalues that decay at least polynomially, with respect to $\P(D,X|G=\EXP)$. Figure~\ref{fig:eigen1} provides evidence of polynomial decay using a product of Gaussian kernels and real Project STAR data ($K_{D^{\EXP}D^{\EXP}}\odot K_{X^{\EXP}X^{\EXP}}$ in Algorithm~\ref{algo:dose}). When using a product of Gaussian kernels with Gaussian data, the eigenfunctions are Hermite polynomials of $d$ and $x$.
\end{remark}

Under these conditions, we arrive at our main result: uniform consistency of long term dose response curves. This result appears to be the first rate in $\sup$ norm for a nonlinear, long term dose response. It accommodates general types of short term rewards, actions, and contexts.

\begin{theorem}[Uniform consistency of long term dose responses]\label{theorem:consistency}
Suppose the conditions of Theorem~\ref{theorem:representation} and Assumptions~\ref{assumption:original},~\ref{assumption:smooth}, and~\ref{assumption:smooth_op} hold. Set $\lambda^{\OBS}=(n^{\OBS})^{-\frac{1}{c^{\OBS}+1/b^{\OBS}}}$ and $\lambda^{\EXP}=(n^{\EXP})^{-\frac{1}{c^{\EXP}+1/b^{\EXP}}}$ in Algorithm~\ref{algo:dose}, which are rate optimal regularization for regression. Then with high probability, 
    \begin{enumerate}
        \item $\|\hat{\theta}-\theta_0\|_{\infty}$, $\|\hat{\theta}^{\EXP}-\theta_0^{\EXP}\|_{\infty}$, and $\|\hat{\theta}^{\OBS}-\theta_0^{\OBS}\|_{\infty}$
         are
        $O \left\{(n^{\OBS})^{-\frac{1}{2}\frac{c^{\OBS}-1}{c^{\OBS}+1/b^{\OBS}}}+ (n^{\EXP})^{-\frac{1}{2}\frac{c^{\EXP}-1}{c^{\EXP}+1/b^{\EXP}}} \right\}$;
        \item $\|\hat{\theta}^{\DS}(\cdot,\tilde{\P})-\theta_0^{\DS}(\cdot,\tilde{\P})\|_{\infty}$ is  $O \left\{(n^{\OBS})^{-\frac{1}{2}\frac{c^{\OBS}-1}{c^{\OBS}+1/b^{\OBS}}}+ (n^{\EXP})^{-\frac{1}{2}\frac{c^{\EXP}-1}{c^{\EXP}+1/b^{\EXP}}} +\tilde{n}^{-1/2} \right\}$.
    \end{enumerate}
\end{theorem}

Appendix~\ref{sec:consistency} gives nonasymptotic error bounds in more detail. The rates reflect the sample sizes $(n^{\OBS},n^{\EXP})$, effective dimension parameters $(b^{\OBS},b^{\EXP})$ and the smoothness parameters $(c^{\OBS},c^{\EXP})$ of the long term regression $\gamma_0^{\OBS}$ and short term embedding $\mu_s^{\EXP}(d,x)$. 
These rates approach $n^{-1/4}$ when $n^{\OBS}=n^{\EXP}$, $(b^{\OBS},b^{\EXP})\rightarrow \infty$, and $(c^{\OBS},c^{\EXP})=2$, i.e. when the sample sizes are of the same order, the effective dimensions are finite, and the targets are smooth. 
Each rate combines minimax optimal rates in RKHS norm for standard nonparametric regression, conditional mean embedding, and unconditional mean embedding \cite{tolstikhin2017minimax,fischer2017sobolev,li2022optimal}. See Section~\ref{sec:assumptions} for translation into mimimax Sobolev norm rates, which are a special case.

\begin{remark}[Compensating for small experiments and complex links]
    Often, experiments are small and long term links are complex. Theorem~\ref{theorem:consistency} demonstrates that a small experiment can be compensated by a simple conditional distribution of short term rewards in the experiment. Meanwhile, a complex link between the short term and the long term can be compensated by a large observational sample.
    These insights distinguish our analysis of kernel methods for long term dose response curves from analyses of kernel methods for other problems.
\end{remark}

We prove uniform rates for an RKHS estimator of the long term dose response curve, under effective dimension and smoothness conditions on the regression and embedding. An interesting direction for future work is whether uniform rate improvements are possible by placing additional assumptions, perhaps building on the techniques from pointwise, mean square, or excess risk analysis \cite{kennedy2017nonparametric,nie2021quasi,foster2023orthogonal,zeng2024continuous}.
\section{Application: Long term dose response of class size}\label{sec:application}

To demonstrate that our proposed kernel methods are practical for empirical research, we evaluate their ability to recover long term dose response curves. Using short term experimental data and long term observational data, our methods measure similar long term effects as an oracle method that has access to long term experimental data. Our methods outperform some benchmarks from previous work that use only long term observational data.

Our exercise directly extends the exercise of \cite{athey2020combining}. Whereas those authors study the long term effect of a binary action (``small'' versus ``large'' classes), we study the long term effect of a continuous action (various class sizes). Figure~\ref{fig:actions} shows that the randomized action $D$ takes values in $\cD=[12, 28]$. It also shows that the supports of the ``small'' and ``large'' classes are overlapping; some ``small'' classes were larger than some ``large'' classes. By modeling $D$ as continuous with a Gaussian kernel $k_{\cD}$, we smoothly share information across different class sizes.

As in previous work, we consider the third grade test score to be the short term reward $S$, and a subsequent test score to be the long term reward $Y$. By choosing different grades as different long term rewards, we evaluate how our methods perform over different horizons. Our variable definitions are identical to \cite{athey2020combining}, except that we use a continuous action.

\begin{figure}
\captionsetup[subfigure]{justification=Centering}
\begin{subfigure}[t]{0.32\textwidth}
         \centering
        \resizebox{\textwidth}{!}{%
       \includegraphics[width=\textwidth]{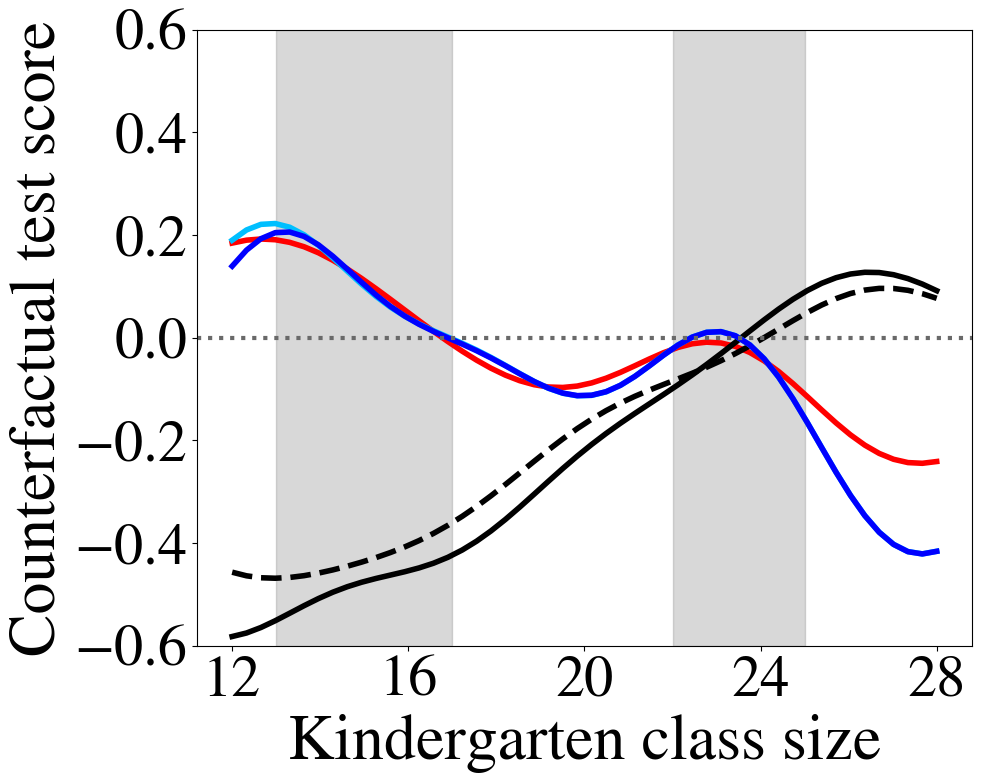}
        }
    \caption{Third grade.}\label{fig:semi3_grey}
\end{subfigure}\hspace{\fill} 
\begin{subfigure}[t]{0.32\textwidth}
          \centering
        \resizebox{\textwidth}{!}{%
      \includegraphics[width=\textwidth]{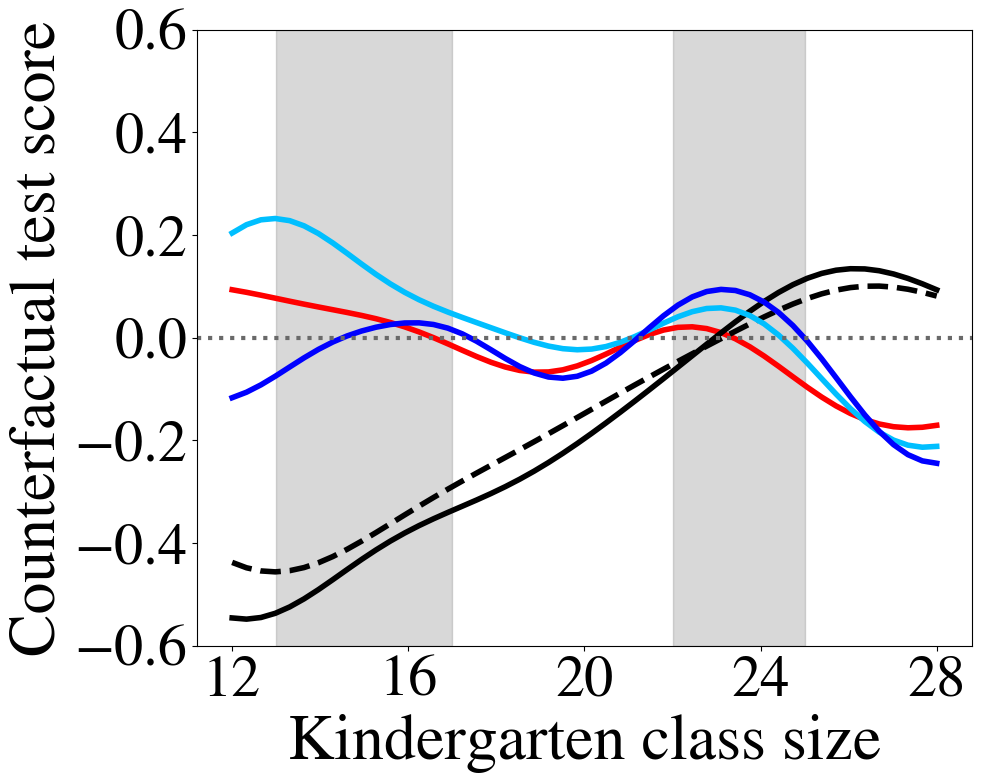}
        }
    \caption{Fourth grade.}\label{fig:semi4_grey}
\end{subfigure}
\hspace{\fill} 
\begin{subfigure}[t]{0.32\textwidth}
          \centering
        \resizebox{\textwidth}{!}{%
      \includegraphics[width=\textwidth]{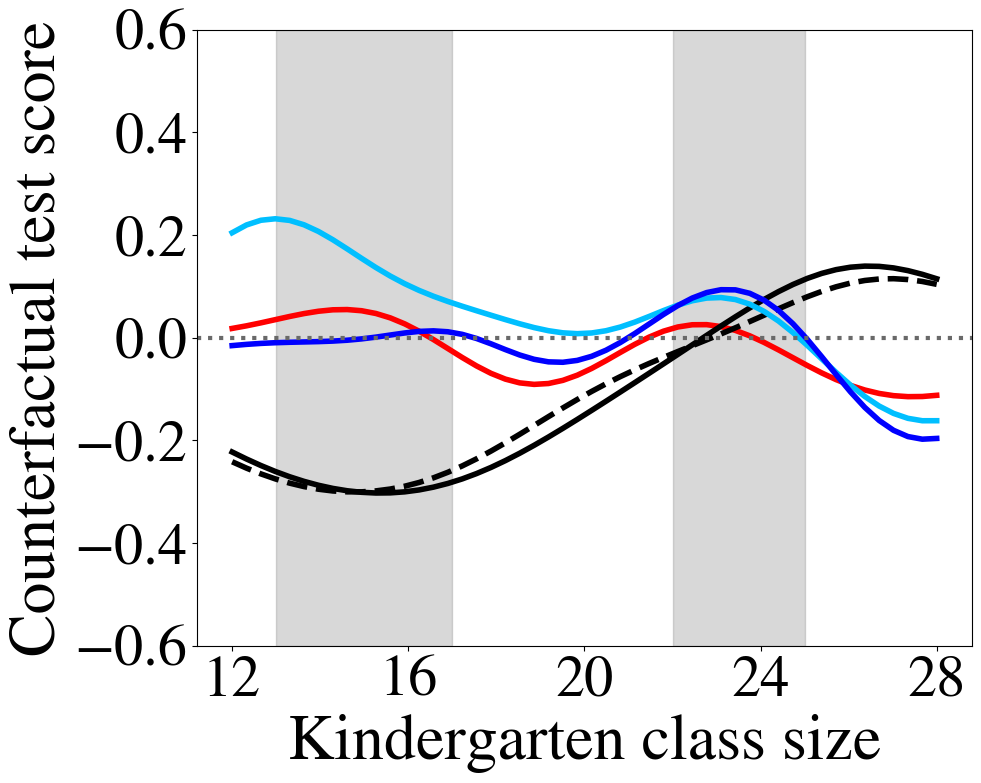}
        }
    \caption{Fifth grade.}\label{fig:semi5_grey}
\end{subfigure}
\caption{Our method recovers the long term dose response of Project STAR over different horizons. 
We compare an oracle (red) with our proposals for the surrogate model (light blue) and the missing-at-random model (dark blue). 
We also visualize benchmarks from previous work, either with (black, dashed) or without (black, solid) covariates. In the background, we indicate ranges of class sizes in the Project STAR protocol (gray).}\label{fig:345_grey}
\end{figure}

The oracle, visualized in red, is estimated from long term experimental data, i.e. joint observations of the randomized action $D$ and long term reward $Y$ in Project STAR. Our goal is to recover similar estimates without access to long term experimental data. Figure~\ref{fig:345_grey} shows that the oracle curve is typically decreasing: larger class sizes appear to cause lower test scores, across horizons. In particular, the oracle estimates are nonlinearly decreasing, from positive counterfactual test scores (above average) to negative counterfactual test scores (below average). As the long term horizon increases, i.e. as the definition of $Y$ corresponds to later grades, the oracle curves flatten: the effect of kindergarten class size on test scores appears to attenuate over time. 

To explore why the curves appear non-monotonic, we visualize in gray the ranges of  class sizes  targeted in the Project STAR protocol. There is a positive bump at $d=22$, which is the beginning of a targeted range, suggesting that there may be selection at this boundary.

The benchmarks, visualized in black, are estimated from joint observations of the nonrandomized action $D$ and long term reward $Y$ in observational data from New York City (NYC) public schools. We consider versions that do (dashed) or do not (solid) adjust for baseline covariates $X$: race, gender, and free and reduced price lunch status. We implement the benchmarks using an estimator from previous work  based on kernel ridge regression \cite{singh2020kernel}. The benchmark estimates are steeply increasing, whereas the oracle estimates are decreasing; a researcher using previous kernel methods for continuous actions would arrive at the opposite of the desired conclusion. 

Finally, we implement our proposals, estimated using short term experimental data from Project STAR and long term observational data from NYC. We consider versions for the surrogate model in Lemma~\ref{lemma:identification} (light blue), and the missing-at-random model in Remark~\ref{remark:additional} (dark blue). The former uses $(D,S)$ from Project STAR and $(S,Y)$ from NYC, while the latter uses $(D,S)$ from Project STAR and $(D,S,Y)$ from NYC. Both versions of our method recover the direction and nonlinear shape of the oracle. The latter version recovers the oracle particularly well, suggesting that the missing-at-random identification may be more applicable than the surrogacy identification in this setting. Appendix~\ref{sec:experiment_details} demonstrates that these trends continue over longer horizons.

\section{Discussion}\label{sec:conclusion}

Long term causal inference is a core challenge in both the development of artificial intelligence and in the evaluation of social programs; exploratory and experimental data are often short term, due to collection expenses, privacy concerns, or ethical considerations. We propose a kernel method that learns a representation of the short term experimental data distribution, in order to extrapolate long term dose response curves. This representation provides interpretable weights. We consider the additional challenge of distribution shift, i.e. of transfer learning.

As a contribution to long term causal inference, we propose what appears to be the first nonlinear, nonparametric estimator for long term dose response curves, by adapting techniques and assumptions from the RKHS statistical framework. We derive a closed form for the estimator using RKHS techniques, and prove that it uniformly consistent across actions under RKHS assumptions. The RKHS framework improves our scientific understanding of real data from Project STAR.

Future work may extend our results to neural network embeddings, or to reinforcement learning problems. Future work may also improve scalability via Nystr\"om approximations.

\spacingset{1}

\bibliographystyle{apalike}

\newpage 

\spacingset{1.9}

\appendix
\section{Extension: Long term counterfactual distributions}\label{sec:distribution}

\subsection{Goal: Distribution of long term reward}

While the mean of counterfactual long term rewards may be of primary interest, additional moments of counterfactual long term rewards may be of secondary interest. For example, the variance and skewness of the long term reward distribution contain important information. In artificial intelligence applications, it is common for the long term reward to be sparse, i.e. to frequently take the value of zero. In this common scenario, the counterfactual distribution of long term rewards is more informative than the counterfactual mean.

For these reasons, we extend our methods and theory: from counterfactual means of long term rewards, to counterfactual distributions of long term rewards.  We propose what appear to be the first nonparametric estimators for long term counterfactual distributions. In our analysis, we relax the condition in Assumption~\ref{assumption:original} that $\cY\subset\R$, to the more general setting where $\cY$ is a Polish space. As such, the long term reward, short term reward, action, and context may be discrete or continuous, and even texts, images, or graphs. The estimands mirror Definition~\ref{def:dose}, replacing $\E(\cdot)$ with $\P(\cdot)$.

\begin{definition}[Long term counterfactual distributions]\label{def:distributions}
   \begin{enumerate}
       \item Counterfactual distribution: $\pi_0(d)=\P\{Y^{(d)}\}$ is the counterfactual long term reward distribution, given action $D=d$, for the average context. 
       \item Distribution shifted counterfactual distribution: $\pi_0^{\DS}(d,\tilde{\P})=\tilde{\P}\{Y^{(d)}\}$ is the counterfactual long term  reward distribution, given action $D=d$, for an alternative context with distribution $\tilde{\P}$. 
       \item Experimental counterfactual distribution: $\pi_0^{\EXP}(d)=\P\{Y^{(d)}|G=\EXP\}$ is the counterfactual long term reward distribution, given action $D=d$, for the average experimental context. 
       \item Observational counterfactual distribution: $\pi_0^{\OBS}(d)=\P\{Y^{(d)}|G=\OBS\}$ is the counterfactual long term reward distribution, given action $D=d$, for the average observational context. 
   \end{enumerate}
\end{definition}

Their identifications are identical to Lemma~\ref{lemma:identification}, replacing $\E(\cdot)$ with $\P(\cdot)$.

\subsection{Method: Short term and long term kernel embeddings}

To embed counterfactual distributions over long term rewards, we define an RKHS $H_{\cY}$ with features $\phi(y)$ and kernel $k_{\cY}(y,y')$.  We impose RKHS regularity conditions, as in Assumption~\ref{assumption:original}.

\begin{assumption}[RKHS regularity]\label{assumption:RKHS_dist}
    \begin{enumerate}
\item The features $\phi(Y)$ are measurable.
        \item The kernel $k_{\cY}$ is bounded: $\sup_{y\in\cY}\|\phi(y)\|_{H_{\cY}}\leq\kappa_y$.
        \item The kernel $k_{\cY}$ is characteristic: for all Borel probability measures $\Q(y)$, the mapping $\Q(y)\mapsto \int \phi(y)\d\Q(y)$ is injective.
    \end{enumerate}
\end{assumption}

Using this long term reward RKHS, we embed the distributions in Definition~\ref{def:distributions}.

\begin{definition}[Long term counterfactual embeddings]\label{def:embeddings}
   \begin{enumerate}
       \item Counterfactual embedding: $\nu_0(d)=\E[\phi \{Y^{(d)}\}]$ is the counterfactual long term reward embedding, given action $D=d$, for the average context. 
       \item Distribution shifted counterfactual embedding: $\nu_0^{\DS}(d,\tilde{\P})=\E_{\tilde{\P}}[\phi\{Y^{(d)}\}]$ is the counterfactual long term  reward embedding, given action $D=d$, for an alternative context with distribution $\tilde{\P}$. 
       \item Experimental counterfactual embedding: $\nu_0^{\EXP}(d)=\E[\phi\{Y^{(d)}\}|G=\EXP]$ is the counterfactual long term reward embedding, given action $D=d$, for the average experimental context. 
       \item Observational counterfactual embedding: $\nu_0^{\OBS}(d)=\E[\phi\{Y^{(d)}\}|G=\OBS]$ is the counterfactual long term reward embedding, given action $D=d$, for the average observational context. 
   \end{enumerate}
\end{definition}

To estimate these embeddings, we replace the long term regression $\gamma_0^{\OBS}(S,X)=\E(Y|S,G=\OBS,X)$ with the long term embedding $\mu_y^{\OBS}(S,X)=\E\{\phi(Y)|S,G=\OBS,X\}$. The identification of the quantities in Definition~\ref{def:embeddings} is identical to Lemma~\ref{lemma:identification}, replacing $\gamma_0^{\OBS}$ with $\mu_y^{\OBS}$.

In the same way that the short term embedding $\mu_s^{\EXP}(d,x)$ has an equivalent short term conditional expectation operator $E_0^{\EXP}$ satisfying $\mu_s^{\EXP}(d,x)=(E_0^{\EXP})^* \phi(d,x)$, the long term embedding $\mu_y^{\OBS}(s,x)$ has an equivalent long term conditional expectation operator $E_0^{\OBS}$ satisfying $\mu_y^{\OBS}(s,x)=(E_0^{\OBS})^* \phi(s,x)$, under analogous regularity conditions.

\begin{theorem}[Representation via short term and long term kernel embeddings]\label{theorem:representation_dist}
  If Assumptions~\ref{assumption:RKHS} and \ref{assumption:RKHS_dist} hold, then the short term kernel embedding $\mu^{\EXP}_s(d,x)=\int \phi(s)\d\P(s|d,G=\EXP,x)$ and the long term kernel embedding 
    $\mu_y^{\OBS}(s,x)=\int \phi(y)\d\P(y|s,G=\OBS,x)$
    exist. If, in addition, the conditions of Lemma~\ref{lemma:identification} hold and $E_0^{\OBS}\in \cL_2(H_{\cY},H_{\cS}\otimes H_{\cX})$, then we have the representations
    \begin{enumerate}
        \item $\nu_0(d)=(E_0^{\OBS})^* \mu_{s,x}(d)$ where $\mu_{s,x}(d)=\int \{\mu_s^{\EXP}(d,x) \otimes \phi(x)\}\d\P(x)$;
        \item  $\nu^{\DS}_0(d)=(E_0^{\OBS})^* \tilde{\mu}_{s,x}(d)$ where $\tilde{\mu}_{s,x}(d)=\int \{\mu_s^{\EXP}(d,x) \otimes \phi(x)\}\d\tilde{\P}(x)$;
        \item $\nu^{\EXP}_0(d)=(E_0^{\OBS})^* \mu'_{s,x}(d)$ where $\mu'_{s,x}(d)=\int \{\mu_s^{\EXP}(d,x) \otimes \phi(x)\}\d\P(x|G=\EXP)$;
        \item $\nu^{\OBS}_0(d)=(E_0^{\OBS})^* \mu''_{s,x}(d)$ where $\mu''_{s,x}(d)=\int \{\mu_s^{\EXP}(d,x) \otimes \phi(x)\}\d\P(x|G=\OBS)$.
    \end{enumerate}
\end{theorem}

These representations generalize those in Theorem~\ref{theorem:representation}. We derive generalized estimators.

\begin{algorithm}[Long term counterfactual embedding estimator]\label{algo:embedding}
As before, let the superscript $\EXP$ or $\OBS$ denote which data are being used. 
Let $n=n^{\OBS}+n^{\EXP}$ be the total number of observations from the distribution $\P$, and let $\tilde{n}$ be the number of observations from the alternative context distribution $\tilde{\P}$.
Let $\odot$ be the elementwise product. The long term counterfactual embedding estimator is
\begin{align*}
    \{\hat{\nu}(d)\}(y)&=\frac{1}{n}\sum_{i=1}^n
K_{yY^{\OBS}}(K_{S^{\OBS}S^{\OBS}}\odot K_{X^{\OBS}X^{\OBS}}+n^{\OBS}\lambda^{\OBS} I^{\OBS})^{-1}\\
&[\{K_{S^{\OBS} S^{\EXP}}(K_{D^{\EXP}D^{\EXP}}\odot K_{X^{\EXP}X^{\EXP}}+n^{\EXP}\lambda^{\EXP}I^{\EXP})^{-1}(K_{D^{\EXP}d}\odot K_{X^{\EXP}X_i})\}\odot K_{X^{\OBS}X_i}],
\end{align*}
where $I^{\OBS}$ 
and $I^{\EXP}$ 
are identity matrices. 
The other estimators are similar. For $\hat{\nu}^{\DS}(d)$, $\hat{\nu}^{\EXP}(d)$, and $\hat{\nu}^{\OBS}(d)$, replace $\frac{1}{n}\sum_{i=1}^n(\cdot)$ with $\frac{1}{\tilde{n}}\sum_{i=1}^{\tilde{n}}(\cdot)$, $\frac{1}{n^{\EXP}}\sum_{i\in \EXP}(\cdot)$, and $\frac{1}{n^{\OBS}}\sum_{i\in \OBS}(\cdot)$, respectively.
\end{algorithm}

Theorem~\ref{theorem:consistency_dist} gives theoretical values of $(\lambda^{\OBS},\lambda^{\EXP})$ to optimally balance bias and variance of the conditional mean embeddings. Appendix~\ref{sec:tuning} gives practical tuning procedures.

Given an estimate of an embedding, a well known procedure called kernel herding allows us to recover the distribution by sampling from it \cite{welling2009herding}.

\begin{algorithm}[Long term counterfactual distribution estimator]\label{algo:distribution}
Given $\hat{\nu}(d)$, sample $(Y_1',Y_2',...)$ from $\pi_0(d)$ recursively via
$
Y_j'=\argmax_{y\in \cY} \left[ \hat{\nu}(d)\}(y) -\frac{1}{j+1}\sum_{\ell=1}^{j-1}k_{\cY}(Y_{\ell}',y)\right].
$
Likewise we can sample from the other long term counterfactual distributions in Definition~\ref{def:distributions}, replacing $\hat{\nu}(d)$ with alternative estimates from Algorithm~\ref{algo:embedding}.
\end{algorithm}

 A histogram of $(Y_j')$ estimates the long term counterfactual density. Empirical moments of $(Y_j')$ estimate the long term counterfactual moments, e.g. the variance and skewness of long term counterfactual rewards.

\subsection{Theory: Weak convergence}

To analyze long term counterfactual embeddings, we extend our previous assumptions for long term dose responses. Importantly, we generalize Assumption~\ref{assumption:smooth} from the long term regression $\gamma_0^{\OBS}$ to the long term conditional expectation operator $E_0^{\OBS}$. We slightly abuse notation, reusing the symbols $(b^{\OBS},c^{\OBS})$ for simplicity.

\begin{assumption}[Original space regularity]\label{assumption:original_dist} 
The long term reward space $\cY$ is Polish space, i.e. separable and completely metrizable.
\end{assumption}

\begin{assumption}[Effective dimension and smoothness for long term conditional expectation operator]\label{assumption:smooth_op_dist}  Eigenvalues decay and the operator  satisfies a source condition:   $\eta^{\OBS}_j(H_{\cS}\otimes H_{\cX})\leq C j^{-(b^{\OBS})}$ and $E_0^{\OBS}\in \cL_2\{H_{\cY},(H_{\cD}\otimes H_{\cX})^{(c^{\OBS})}\}$ for some $b^{\OBS}\geq 1$ and $c^{\OBS}\in(1,2]$.
\end{assumption}

\begin{remark}[Interpretation and diagnostic]\label{remark:dist}
    We assume that the long term generalized regression $\mu_y^{\OBS}(s,x)=\E\{\phi(Y)|S=s,G=\OBS,X=x\}=(E_0^{\OBS})^*\phi(s,x)$ is smooth relative to the eigenfunctions of the product kernel $k_{\cS}(s,s')k_{\cX}(x,x')$, and that this kernel has eigenvalues that decay at least polynomially, with respect to $\P(S,X|G=\OBS)$. Figure~\ref{fig:eigen2} provides evidence of polynomial decay using a product of Gaussian kernels and real Project STAR data ($K_{S^{\OBS}S^{\OBS}}\odot K_{X^{\OBS}X^{\OBS}}$ in Algorithm~\ref{algo:embedding}). When using a product of Gaussian kernels with Gaussian data, the eigenfunctions are Hermite polynomials of $s$ and $x$.
\end{remark}

\begin{theorem}[Uniform consistency of long term counterfactual embeddings]\label{theorem:consistency_dist}
Suppose the conditions of Theorem~\ref{theorem:representation_dist} and Assumptions~\ref{assumption:original},~\ref{assumption:smooth_op},~\ref{assumption:original_dist}, and~\ref{assumption:smooth_op_dist} hold. Set $\lambda^{\OBS}=(n^{\OBS})^{-\frac{1}{c^{\OBS}+1/b^{\OBS}}}$ and $\lambda^{\EXP}=(n^{\EXP})^{-\frac{1}{c^{\EXP}+1/b^{\EXP}}}$ in Algorithm~\ref{algo:distribution}, which are rate optimal regularization for conditional mean embedding. Then with high probability, 
    \begin{enumerate}
        \item $\sup_{d\in\cD}\|\hat{\nu}(d)-\nu_0(d)\|_{H_{\cY}}$, $\sup_{d\in\cD}\|\hat{\nu}^{\EXP}(d)-\nu_0^{\EXP}(d)\|_{H_{\cY}}$, and $\sup_{d\in\cD}\|\hat{\nu}^{\OBS}(d)-\nu_0^{\OBS}(d)\|_{H_{\cY}}$
         are
        $O \left\{(n^{\OBS})^{-\frac{1}{2}\frac{c^{\OBS}-1}{c^{\OBS}+1/b^{\OBS}}}+ (n^{\EXP})^{-\frac{1}{2}\frac{c^{\EXP}-1}{c^{\EXP}+1/b^{\EXP}}} \right\}$;
        \item $\sup_{d\in\cD}\|\hat{\nu}^{\DS}(d,\tilde{\P})-\nu_0^{\DS}(d,\tilde{\P})\|_{H_{\cY}}$ is  $O \left\{(n^{\OBS})^{-\frac{1}{2}\frac{c^{\OBS}-1}{c^{\OBS}+1/b^{\OBS}}}+ (n^{\EXP})^{-\frac{1}{2}\frac{c^{\EXP}-1}{c^{\EXP}+1/b^{\EXP}}} +\tilde{n}^{-1/2} \right\}$.
    \end{enumerate}
\end{theorem}

Finally, we prove that the random variables $(Y_j')$, sampled from the long term counterfactual embedding estimate, weakly converge to the desired long term counterfactual distribution.

\begin{assumption}[Further regularity]\label{assumption:further}
    \begin{enumerate}
   \item The RKHS $H_{\cY}$ is contained in $\cC$, the space of bounded, continuous, real valued functions that vanish at infinity.
   \item The original space $\cY$ is locally compact.
    \end{enumerate}
\end{assumption}

For example, $\cY=\mathbb{R}^p$ is both Polish and locally compact. 

\begin{corollary}[Weak convergence for long term counterfactual distributions]\label{cor:dist}
    Suppose the conditions of Theorem~\ref{theorem:consistency_dist} and Assumption~\ref{assumption:further} hold. Suppose $(Y_j')$ are sampled according to Algorithm~\ref{algo:distribution}. Then $(Y_j')\rightsquigarrow \pi_0(d)$. The same result holds for other counterfactual distributions in Definition~\ref{def:distributions}, replacing $\hat{\nu}(d)$ in Algorithm~\ref{algo:distribution} with alternative estimates from Algorithm~\ref{algo:embedding}.
\end{corollary}

\section{Identification}\label{sec:identification}

We state the identifying assumptions for Lemma~\ref{lemma:identification}. Assumption~\ref{assumption:surrogate} is well known. Assumption~\ref{assumption:shift} is a modest extension for transfer learning. The same arguments identify long term counterfactual distributions in Appendix~\ref{sec:distribution}, replacing $\E(\cdot)$ with $\P(\cdot)$ in the proofs.

\begin{assumption}[Surrogate model]\label{assumption:surrogate}
We recap well known assumptions \cite{prentice1989surrogate,athey2020estimating}.
    \begin{enumerate}
        \item Causal consistency: if $D=d$ then $S=S^{(d)}$ and $Y=Y^{(d)}$.
        \item Unconfounded selection: $G\indep Y^{(d)},S^{(d)} |X$.
        \item Selection overlap: if $\d\P(x)>0$ then $\P(G=\EXP|x)>0$; and if $\d\P(s,x)>0$ then $\P(G=\OBS|s,x)>0$.
        \item Experimental action: $D\indep Y^{(d)},S^{(d)}|G=\EXP,X$.
        \item Experimental action overlap: if $\d\P(G=\EXP,x)>0$ then $\d\P(d|G=\EXP,x)>0$.
        \item Surrogacy: $D\indep Y|S,G=\EXP,X$.
        \item Comparability: $G\indep Y|S,X$.
    \end{enumerate}
\end{assumption}

\begin{assumption}[Surrogate model distribution shift]\label{assumption:shift}
We introduce the following assumptions.
    \begin{enumerate}
        \item Experimental transferability: $\tilde{\P}(S,D,G=\EXP,X)=\P(S|D,G=\EXP,X)\tilde{\P}(D,G=\EXP,X)$.
        \item Experimental overlap: $\tilde{\P}(D,G=\EXP,X)$ is absolutely continuous with respect to $\P(D,G=\EXP,X)$.
        \item Observational transferability: $\tilde{\P}(Y,S,G=\OBS,X)=\P(Y|S,G=\OBS,X)\tilde{\P}(S,G=\OBS,X)$
        \item Observational overlap: $\tilde{\P}(S,G=\OBS,X)$ is absolutely continuous with respect to $\P(S,G=\OBS,X)$.
    \end{enumerate}
\end{assumption}

\begin{proof}[Proof of Lemma~\ref{lemma:identification}]
    For completeness, we present the known proof. The distribution shift estimand is a modest extension.
    To begin, write
    \begin{align*}
        \E\{Y^{(d)}|X\}&=\E\{Y^{(d)}|G=\EXP,X\} 
        =\E\{Y^{(d)}|D=d,G=\EXP,X\} 
          =\E(Y|D=d,G=\EXP,X) \\
          &=\E\{\E(Y|S,D=d,G=\EXP,X)|D=d,G=\EXP,X \}
    \end{align*}
    by unconfounded selection, experimental action, causal consistency, and the law of iterated expectations.
    Focusing on the inner expectation,
    \begin{align*}
        \E(Y|S,D=d,G=\EXP,X)
        &= \E(Y|S,G=\EXP,X)
        =\E(Y|S,G=\OBS,X)
    \end{align*}
    by surrogacy and comparability. In summary,
    $$
    \E\{Y^{(d)}|X\}=\int \E(Y|s,G=\OBS,X) \d\P(s|D=d,G=\EXP,X).
    $$
    By the law of iterated expectations,
   $$
   \E\{Y^{(d)}\}=\E[\E\{Y^{(d)}|X\}],\quad \E_{\tilde{\P}}\{Y^{(d)}\}=\E_{\tilde{\P}}[\E_{\tilde{\P}}\{Y^{(d)}|X\}],\quad \E\{Y^{(d)}|G\}=\E[\E\{Y^{(d)}|G,X\}|G].
   $$
    This completes the argument for $\theta_0$. For $\theta_0^{\DS}$, experimental transferability and observational transferability within Assumption~\ref{assumption:shift} imply that $\P(S|D,G=\EXP,X)$ and $\P(Y|S,G=\OBS,X)$ are invariant across $\P$ and $\tilde{\P}$. Finally, for $\theta_0^{\EXP}$ and $\theta_0^{\OBS}$, unconfounded selection implies $\E\{Y^{(d)}|G,X\}=\E\{Y^{(d)}|X\}$.
\end{proof}
\section{Algorithm derivation}\label{sec:algo}

\subsection{Short term kernel embedding}

\begin{proof}[Proof of Theorem~\ref{theorem:representation}]
    Assumption~\ref{assumption:RKHS} implies Bochner integrability \cite[Definition A.5.20]{steinwart2008support}, which permits the exchange of the expectation and inner product. Therefore by Lemma~\ref{lemma:identification}, $ \theta_0(d)$ equals
    \begin{align*}
       &\int \int \gamma_0^{\OBS}(s,x)\d\P(s|d,G=\EXP,x)\d\P(x)
       =\int \int \left\langle \gamma^{\OBS}_0, \phi(s)\otimes \phi(x) \right\rangle_{H}\d\P(s|d,G=\EXP,x)\d\P(x) \\
        &=\int  \left\langle \gamma^{\OBS}_0, \int\phi(s)\d\P(s|d,G=\EXP,x) \otimes \phi(x) \right\rangle_{H}\d\P(x)
        =\int  \left\langle \gamma^{\OBS}_0, \mu_s^{\EXP}(d,x) \otimes \phi(x) \right\rangle_{H}\d\P(x) \\
        &=\left\langle \gamma^{\OBS}_0,\int   \mu_s^{\EXP}(d,x) \otimes \phi(x) \d\P(x) \right\rangle_{H}
        =\left\langle \gamma^{\OBS}_0,\mu_{s,x}(d) \right\rangle_{H}.
    \end{align*}
    The arguments for the other long term dose responses are similar, replacing $\P(x)$ with other distributions, according to Lemma~\ref{lemma:identification}.
\end{proof}

\begin{proof}[Proof of Theorem~\ref{theorem:representation_dist}]
    Assumptions~\ref{assumption:RKHS} and~\ref{assumption:RKHS_dist} imply Bochner integrability \cite[Definition A.5.20]{steinwart2008support}, which permits the exchange of the expectation and inner product. Therefore, $ \nu_0(d)$ equals
    \begin{align*}
       &\int \int \mu_y^{\OBS}(s,x)\d\P(s|d,G=\EXP,x)\d\P(x)
       =\int \int (E_0^{\OBS})^* \{\phi(s)\otimes \phi(x)\} \d\P(s|d,G=\EXP,x)\d\P(x) \\
        &=\int  (E_0^{\OBS})^* \left\{\int\phi(s)\d\P(s|d,G=\EXP,x) \otimes \phi(x)\right\} \d\P(x)
        =\int  (E_0^{\OBS})^* \left\{\mu_s^{\EXP}(d,x) \otimes \phi(x)\right\} \d\P(x) \\
        &=(E_0^{\OBS})^*\left\{\int   \mu_s^{\EXP}(d,x) \otimes \phi(x) \d\P(x)\right\} 
        =(E_0^{\OBS})^*\left\{\mu_{s,x}(d)\right\}.
    \end{align*}
    The arguments for the other long term counterfactual embeddings are similar, replacing $\P(x)$ with other distributions, as in the proof of Theorem~\ref{theorem:representation}.
\end{proof}

\subsection{Closed form solution}

\begin{proof}[Derivation of Algorithm~\ref{algo:dose}]
    By standard arguments \cite{kimeldorf1971some}, 
    $$
    \hat{\gamma}^{\OBS}(\cdot,x)=(Y^{\OBS})^{\top}(K_{S^{\OBS}S^{\OBS}}\odot K_{X^{\OBS}X^{\OBS}}+n^{\OBS}\lambda^{\OBS} I^{\OBS})^{-1}(K_{S^{\OBS}(\cdot)}\odot K_{X^{\OBS}x}).
    $$
    By \cite[Algorithm 1]{singh2019kernel},
    $$
    \{\hat{\mu}_s^{\EXP}(d,x)\}(\cdot)=K_{(\cdot) S^{\EXP}}(K_{D^{\EXP}D^{\EXP}}\odot K_{X^{\EXP}X^{\EXP}}+n^{\EXP}\lambda^{\EXP}I^{\EXP})^{-1}(K_{D^{\EXP}d}\odot K_{X^{\EXP}x}).
    $$
    Matching empty arguments gives the closed form for $\left\langle \hat{\gamma}^{\OBS}_0, \hat{\mu}_s^{\EXP}(d,x) \otimes \phi(x) \right\rangle_{H}$ for placeholder $x$. 

    Since $\theta_0(d)=\int  \left\langle \gamma^{\OBS}_0, \mu_s^{\EXP}(d,x) \otimes \phi(x) \right\rangle_{H}\d\P(x)$, analogously we have the expression $\hat{\theta}(d)=n^{-1}\sum_{i=1}^n \left\langle \gamma^{\OBS}_0, \mu_s^{\EXP}(d,X_i) \otimes \phi(X_i) \right\rangle_{H}$, which completes the derivation for $\hat{\theta}(d)$.

    The arguments for the other long term dose responses are similar, replacing $n^{-1}\sum_{i=1}^n(\cdot)$ with other empirical distributions, according to Lemma~\ref{lemma:identification}.    
\end{proof}

\begin{proof}[Derivation of Algorithm~\ref{algo:embedding}]
    By \cite[Algorithm 1]{singh2019kernel},
      $$
    \{\hat{\mu}_y^{\OBS}(\cdot,x)\}(y)=K_{yY^{\OBS}}(K_{S^{\OBS}S^{\OBS}}\odot K_{X^{\OBS}X^{\OBS}}+n^{\OBS}\lambda^{\OBS})^{-1}(K_{S^{\OBS}(\cdot)}\odot K_{X^{\OBS}x}).
    $$
    The derivation then mirrors Algorithm~\ref{algo:embedding}, replacing $\hat{\gamma}^{\OBS}(\cdot,x)$ with $\{\hat{\mu}_y^{\OBS}(\cdot,x)\}(y)$.
\end{proof}

\subsection{Long term reward weights}

\begin{proof}[Proof of Corollary~\ref{cor:weights}]
    By Algorithm~\ref{algo:dose}, $\hat{\theta}(d)=\frac{1}{n}\sum_{i=1}^n
(Y^{\OBS})^{\top}\beta_i(d)$, where $\beta_i(d)\in\R^{n^{\OBS}}$ is
\begin{align*}
    \beta_i(d)&=(K_{S^{\OBS}S^{\OBS}}\odot K_{X^{\OBS}X^{\OBS}}+n^{\OBS}\lambda^{\OBS} I^{\OBS})^{-1}\\
&[\{K_{S^{\OBS} S^{\EXP}}(K_{D^{\EXP}D^{\EXP}}\odot K_{X^{\EXP}X^{\EXP}}+n^{\EXP}\lambda^{\EXP}I^{\EXP})^{-1}(K_{D^{\EXP}d}\odot K_{X^{\EXP}X_i})\}\odot K_{X^{\OBS}X_i}].
\end{align*}
Denote by $\beta_{ij}(d)\in\R $ the $j$th entry of $\beta_i(d)\in\R^{n^{\OBS}}$. Then $\hat{\alpha}_j(d)=\frac{n^{\OBS}}{n}\sum_{i=1}^n \beta_{ij}(d)$  since
\begin{align*}
    \hat{\theta}(d)
    &=\frac{1}{n}\sum_{i=1}^n
(Y^{\OBS})^{\top}\beta_i(d) 
=\frac{1}{n}\sum_{i=1}^n \sum_{j\in\OBS}
Y_j \beta_{ij}(d) 
=\frac{1}{n^{\OBS}} \sum_{j\in\OBS}
Y_j \frac{n^{\OBS}}{n}\sum_{i=1}^n \beta_{ij}(d) 
=\frac{1}{n^{\OBS}} \sum_{j\in\OBS}
Y_j \hat{\alpha}_j(d).
\end{align*}
\end{proof}
\section{Hyperparameter tuning}\label{sec:tuning}

\subsection{Kernels}

Consider the real valued kernel $k_{\cA}(a,a')$ over $\cA$. Throughout, we use the popular Gaussian kernel $k_{\cA}(a,a')=\exp\left(-\frac{1}{2}\frac{\|a-a'\|_{\cA}^2}{\iota^2}\right)$, which satisfies Assumption~\ref{assumption:RKHS}. Here, $\iota$ is a hyperparameter called the lengthscale, which we tune using the well known median heuristic: we set $\iota$ equal to the median interpoint distance of $(A_i)$, where the interpoint distance between $A_i$ and $A_j$ is $\|A_i-A_j\|_{\cA}$.

When $\cA$ is multidimensional, we use the product of kernels for each dimension. For example, if $\cA\subset\R^p$, we take $k(a,a')=\prod_{j=1}^p \exp\left\{-\frac{1}{2}\frac{(a_j-a_j')^2}{\iota_j^2}\right\}$. We set each lengthscale according to the median interpoint distance for that dimension.

\subsection{Long term regression regularization}

Consider the kernel ridge regression of $Y\in\cY\subset \R$ on $\phi(A)\in H_{\cA}$ with regularization $\lambda$:
$$
\hat{f}=\argmin_{f\in H_{\cA}} \frac{1}{n}\sum_{i=1}^n \{Y_i-\langle f,\phi(A_i) \rangle_{H_{\cA}}\}^2+\lambda\|f\|_{H_{\cA}}^2.
$$
This is a simplified version of the long term regression estimator $\hat{\gamma}^{\OBS}$. We recap known tuning procedures with closed form solutions: leave-one-out cross validation (LOOCV), and generalized cross validation (GCV). Both are similar in practice. The latter is asymptotically optimal for regression in both $\L^2$ and $H_{\cA}$ norms, so it is compatible with our theoretical analysis. Both procedures are well known; see e.g. \cite{singh2020kernel} for their derivation from first principles.

\begin{algorithm}[LOOCV for kernel ridge regression]\label{algo:cv}
    Construct the $\R^{n\times n}$ matrices $C_{\lambda}=I-K_{AA}(K_{AA}+n\lambda I)^{-1}$ and $\tilde{C}_{\lambda}=\text{diag}(C_{\lambda})$, where $\tilde{C}_{\lambda}$ has the same diagonal entries as $C_{\lambda}$ and off diagonal entries of zero. Set $\lambda^*=\argmin_{\lambda\in \Lambda} n^{-1}\|\tilde{C}_{\lambda}^{-1}C_{\lambda} Y\|_2^2$.
\end{algorithm}

\begin{algorithm}[GCV for kernel ridge regression]\label{algo:gcv}
    Set $\lambda^*=\argmin_{\lambda\in \Lambda} n^{-1}\|\{\tr(C_{\lambda})\}^{-1}C_{\lambda} Y\|_2^2$.
\end{algorithm}

\subsection{Short term conditional expectation operator regularization}

Consider the generalized kernel ridge regression of $\phi(A) \in H_{\cA}$ on $\phi(B)\in H_{\cB}$ with regularization $\lambda$:
$$
\hat{E}
        =\argmin_{E\in \cL_2(H_{\cA},H_{\cB})} \frac{1}{n} \sum_{i=1}^n\{\phi(A_i)-E^*
        \phi(B_i)\}^2+\lambda \|E\|_{\cL_2(H_{\cA},\otimes H_{\cB})}^2.
$$
This is a simplified version of the short term conditional expectation operator estimator $\hat{E}^{\EXP}$. We recap leave-one-out cross validation (LOOCV) and generalized cross validation (GCV). Their derivations from first principles and their properties are similar to Algorithms~\ref{algo:cv} and~\ref{algo:gcv} above.

\begin{algorithm}[LOOCV for conditional expectation operator]\label{algo:cv_op}
    Construct the $\R^{n\times n}$ matrices:
    \begin{enumerate}
        \item $R=K_{BB}(K_{BB}+n\lambda I)^{-1}$;
        \item $S$ with $S_{ij}=1(i=j)\left(\frac{1}{1-R_{ii}}\right)^2$ where $R_{ii}$ is the $i$th diagonal element of $R$;
        \item $T=S(K_{AA}-2K_{AA}R^{\top}+RK_{AA}R^{\top})$.
    \end{enumerate}
   Set $\lambda^*=\argmin_{\lambda\in \Lambda} n^{-1}\tr(T)$.
\end{algorithm}

\begin{algorithm}[GCV for conditional expectation operator]
   Construct the $\R^{n\times n}$ matrices:
    \begin{enumerate}
        \item $R=K_{BB}(K_{BB}+n\lambda I)^{-1}$;
        \item $S=\{\tr(I-R)\}^{-2}I$;
        \item $T=S(K_{AA}-2K_{AA}R^{\top}+RK_{AA}R^{\top})$.
    \end{enumerate}
   Set $\lambda^*=\argmin_{\lambda\in \Lambda} n^{-1}\tr(T)$.
\end{algorithm}

\section{Uniform consistency proof}\label{sec:consistency}

\subsection{Nonasymptotic rates from the literature}

To lighten notation, let $a\lesssim b$ mean that there exists an absolute constant $C<\infty$ such that $a\leq C b$. Consider the abstract regression $f_0(\cdot)=\E\{Y|A=(\cdot)\}$ estimated from $n$ i.i.d. observations $(Y_i,A_i)$ via the kernel ridge regression estimator $\hat{f}(\cdot)$.

\begin{lemma}[Kernel ridge regression rate; e.g. Theorem 1 of \cite{fischer2017sobolev}]\label{lemma:regression}
    Suppose the output $Y$ and features $\phi(A)$ are measurable and bounded; and the original space $\cY\subset\R$ while $\cA$ is Polish. Suppose that eigenvalues decay and the regression satisfies a source condition: $\eta_j(H_{\cA})\lesssim j^{-b_0}$ and $f_0\in H^{c_0}_{\cA}$, for some $b_0\geq 1$ and $c_0\in(1,2]$. Then there exists an $n_0 \in\N$ such that for all $n\geq n_0$, with probability $1-\delta$, 
    $$
    \|\hat{f}-f_0\|_{H_{\cA}}\lesssim r(\delta,n,b_0,c_0)=\ln(4/\delta)n^{-\frac{1}{2}\frac{c_0-1}{c_0+1/b_0}}.
    $$
\end{lemma}

See \cite[Remark S16]{singh2020kernel} for elaboration on $n_0$.

Next, consider the abstract mean embedding $\mu_a=\int \phi(a)\d\P(a)$ estimated from $n$ i.i.d. observations $(A_i)$ via the average $\hat{\mu}_a=\frac{1}{n}\sum_{i=1}^n \phi(A_i)$.

\begin{lemma}[Unconditional mean embedding rate; Theorem 15 of \cite{altun2006unifying}]\label{lemma:unconditional}
    Suppose the features $\phi(A)$ are measurable and bounded; the kernel $k_{\cA}$ is characteristic; and the original space $\cA$ is Polish. Then with probability $1-\delta$,  
    $$
    \|\hat{\mu}_a-\mu_a\|_{H_{\cA}}\lesssim r_{\mu}(\delta,n)= \frac{\ln(2/\delta)}{n^{1/2}}.
    $$
\end{lemma}

Finally, consider the abstract conditional expectation operator $E_0: H_{\cA}\rightarrow H_{\cB}$ satisfying $f(\cdot)\mapsto \E\{f(A)|B=(\cdot)\}$, with the associated conditional mean embedding $\mu_a(b)=E_0^* \{\phi(b)\}$. It is estimated from $n$ i.i.d. observations $(A_i,B_i)$ via the generalized kernel ridge regression estimator
$$
 \hat{E}
        =\argmin_{E\in \cL_2(H_{\cA},H_{\cB})} \frac{1}{n} \sum_{i=1}^n[\phi(A_i)-E^*
        \{\phi(B_i)\} ]^2+\lambda  \|E\|_{\cL_2(H_{\cA},H_{\cB})}^2.
$$

\begin{lemma}[Conditional mean embedding rate; Proposition S5 of \cite{singh2020kernel}]\label{lemma:conditional}
    Suppose the features $\phi(A)$ and $\phi(B)$ are measurable and bounded; the kernel $k_{\cA}$ is characteristic; and the original spaces $\cA$ and $\cB$ are Polish. Suppose that eigenvalues decay and the operator satisfies a source condition: $\eta_j(H_{\cB})\lesssim j^{-b_0}$ and $E_0\in \cL_2(H_{\cA},H^{c_0}_{\cB})$, for some $b_0\geq 1$ and $c_0\in(1,2]$. Then there exists an $n_0 \in\N$ such that for all $n\geq n_0$, with probability $1-\delta$, 
    $$
    \sup_{b\in \cB}\|\hat{\mu}_a(b)-\mu_a(b)\|_{H_{\cA}}\lesssim \|\hat{E}-E_0\|_{\cL_2(H_{\cA},H_{\cB})}\lesssim r_E(\delta,n,b_0,c_0)=\ln(4/\delta)n^{-\frac{1}{2}\frac{c_0-1}{c_0+1/b_0}}.
    $$
\end{lemma}

As before, see \cite[Remark S16]{singh2020kernel} for elaboration on $n_0$.

\subsection{Main results}

To lighten notation, define
\begin{align*}
    \Delta(d)&=\frac{1}{n}\sum_{i=1}^n \left\{ \hat{\mu}_s^{\EXP}(d,X_i)\otimes \phi(X_i) \right\}-\int  \left\{ \mu_s^{\EXP}(d,x)\otimes \phi(x) \right\} \d\P(x) \\
     \Delta^{\DS}(d)&=\frac{1}{\tilde{n}}\sum_{i=1}^{\tilde{n}} \left\{ \hat{\mu}_s^{\EXP}(d,\tilde{X}_i)\otimes \phi(\tilde{X}_i) \right\}-\int  \left\{ \mu_s^{\EXP}(d,x)\otimes \phi(x) \right\} \d\tilde{\P}(x) \\
      \Delta^{\EXP}(d)&=\frac{1}{n^{\EXP}}\sum_{i \in\EXP} \left\{ \hat{\mu}_s^{\EXP}(d,X_i)\otimes \phi(X_i) \right\}-\int  \left\{ \mu_s^{\EXP}(d,x)\otimes \phi(x) \right\} \d\P(x|G=\EXP) \\
       \Delta^{\OBS}(d)&=\frac{1}{n^{\OBS}}\sum_{i\in\OBS} \left\{ \hat{\mu}_s^{\EXP}(d,X_i)\otimes \phi(X_i) \right\}-\int  \left\{ \mu_s^{\EXP}(d,x)\otimes \phi(x) \right\} \d\P(x|G=\OBS).
\end{align*}

\begin{proposition}[Fused embedding rate]\label{prop:rate}
    Suppose Assumptions~\ref{assumption:RKHS},~\ref{assumption:original}, and~\ref{assumption:smooth_op} hold. The following statements hold for $n^{\EXP}$ sufficiently large.
    \begin{enumerate}
        \item With probability $1-2\delta$, $\sup_{d\in\cD}\|\Delta(d)\|_{H_{\cS}\otimes H_{\cX}}\lesssim r_E(\delta,n^{\EXP},b^{\EXP},c^{\EXP})+r_{\mu}(\delta,n)$.
        \item With probability $1-2\delta$, $\sup_{d\in\cD}\|\Delta^{\DS}(d)\|_{H_{\cS}\otimes H_{\cX}}\lesssim r_E(\delta,n^{\EXP},b^{\EXP},c^{\EXP})+r_{\mu}(\delta,\tilde{n})$.
        \item With probability $1-2\delta$, $\sup_{d\in\cD}\|\Delta^{\EXP}(d)\|_{H_{\cS}\otimes H_{\cX}}\lesssim r_E(\delta,n^{\EXP},b^{\EXP},c^{\EXP})+r_{\mu}(\delta,n^{\EXP})$.
        \item With probability $1-2\delta$, $\sup_{d\in\cD}\|\Delta^{\OBS}(d)\|_{H_{\cS}\otimes H_{\cX}}\lesssim r_E(\delta,n^{\EXP},b^{\EXP},c^{\EXP})+r_{\mu}(\delta,n^{\OBS})$.
    \end{enumerate}
\end{proposition}

\begin{proof}
    We prove the result for $\Delta(d)$; the other results are similar. By the triangle inequality,
    \begin{align*}
        \sup_{d\in\cD}\|\Delta(d)\|_{H_{\cS}\otimes H_{\cX}}&\leq 
  \sup_{d\in\cD}\left\|\frac{1}{n}\sum_{i=1}^n \left\{ \hat{\mu}_s^{\EXP}(d,X_i)\otimes \phi(X_i) \right\}-\frac{1}{n}\sum_{i=1}^n\left\{ \mu_s^{\EXP}(d,X_i)\otimes \phi(X_i) \right\}\right\|_{H_{\cS}\otimes H_{\cX}}
\\
&+
  \sup_{d\in\cD}\left\|\frac{1}{n}\sum_{i=1}^n\left\{ \mu_s^{\EXP}(d,X_i)\otimes \phi(X_i) \right\}-\int  \left\{ \mu_s^{\EXP}(d,x)\otimes \phi(x) \right\} \d\P(x)
        \right\|_{H_{\cS}\otimes H_{\cX}}.
    \end{align*}
    
    Within the former term, the norm equals
    \begin{align*}
        &\left\|\frac{1}{n}\sum_{i=1}^n \left[\left\{ \hat{\mu}_s^{\EXP}(d,X_i)-\mu_s^{\EXP}(d,X_i) \right\}\otimes \phi(X_i)\right]\right\|_{H_{\cS}\otimes H_{\cX}}
        \leq \frac{1}{n}\sum_{i=1}^n \left\|\left\{ \hat{\mu}_s^{\EXP}(d,X_i)-\mu_s^{\EXP}(d,X_i) \right\}\otimes \phi(X_i)\right\|_{H_{\cS}\otimes H_{\cX}} \\
        &=\frac{1}{n}\sum_{i=1}^n \left\| \hat{\mu}_s^{\EXP}(d,X_i)-\mu_s^{\EXP}(d,X_i) \right\|_{H_{\cS}} \cdot \left\| \phi(X_i)\right\|_{H_{\cX}}
        \leq \sup_{ x\in\cX} \left\| \hat{\mu}_s^{\EXP}(d,x)-\mu_s^{\EXP}(d,x) \right\|_{H_{\cS}} \cdot \sup_{x\in X} \left\| \phi(x)\right\|_{H_{\cX}}
   \end{align*}
by the triangle inequality, definition of tensor product norm, and definition of supremum. Therefore by boundedness of the kernel and Lemma~\ref{lemma:conditional}, for $n^{\EXP}$ sufficiently large, with probability $1-\delta$, the former term is of order $r_E(\delta,n^{\EXP},b^{\EXP},c^{\EXP}).
$

By Lemma~\ref{lemma:unconditional}, with probability $1-\delta$, the latter term is of order $r_{\mu}(\delta,n)$.   
\end{proof}

\begin{proof}[Proof of Theorem~\ref{theorem:consistency}]
    Fix $d\in\cD$. By Theorem~\ref{theorem:representation}, 
    \begin{align*}
        \hat{\theta}(d)-\theta_0(d)
        &=\left\langle \hat{\gamma}^{\OBS}, \frac{1}{n}\sum_{i=1}^n \{\hat{\mu}_s^{\EXP}(d,X_i)\otimes \phi(X_i)\}\right\rangle_H
        - \left\langle \gamma_0^{\OBS}, \int \{\mu_s^{\EXP}(d,x)\otimes \phi(x)\}\d\P(x)\right\rangle_H \\
        &=\left\langle \hat{\gamma}^{\OBS}, \Delta(d) \right\rangle_H
        + \left\langle \hat{\gamma}^{\OBS}-\gamma_0^{\OBS}, \int \{\mu_s^{\EXP}(d,x)\otimes \phi(x)\}\d\P(x)\right\rangle_H \\
        &=\left\langle \hat{\gamma}^{\OBS}-\gamma_0^{\OBS}, \Delta(d) \right\rangle_H
        +\left\langle\gamma_0^{\OBS}, \Delta(d) \right\rangle_H
        + \left\langle \hat{\gamma}^{\OBS}-\gamma_0^{\OBS}, \int \{\mu_s^{\EXP}(d,x)\otimes \phi(x)\}\d\P(x)\right\rangle_H.
    \end{align*}
    Therefore, by the triangle and Cauchy-Schwarz inequalities,
\begin{align*}
    &|\hat{\theta}(d)-\theta_0(d)|
    \leq \left|\left\langle \hat{\gamma}^{\OBS}-\gamma_0^{\OBS}, \Delta(d) \right\rangle_H\right|
        +\left|\left\langle\gamma_0^{\OBS}, \Delta(d) \right\rangle_H\right|
        + \left|\left\langle \hat{\gamma}^{\OBS}-\gamma_0^{\OBS}, \int \{\mu_s^{\EXP}(d,x)\otimes \phi(x)\}\d\P(x)\right\rangle_H\right| \\
        &\leq \|\hat{\gamma}^{\OBS}-\gamma_0^{\OBS}\|_{H} \cdot \|\Delta(d)\|_H
        +\|\gamma_0^{\OBS}\|_H\cdot\|\Delta(d) \|_H
        + \|\hat{\gamma}^{\OBS}-\gamma_0^{\OBS}\|_H \cdot \left\| \int \{\mu_s^{\EXP}(d,x)\otimes \phi(x)\}\d\P(x)\right\|_H.
\end{align*}

By Lemma~\ref{lemma:regression}, for $n^{\OBS}$ sufficiently large, with probability $1-\delta$,  $\|\hat{\gamma}^{\OBS}-\gamma_0^{\OBS}\|_H \lesssim r(\delta,n^{\OBS},b^{\OBS},c^{\OBS}).$ By Proposition~\ref{prop:rate}, for $n^{\EXP}$ sufficiently large, with probability $1-2\delta$, $\sup_{d\in\cD}\|\Delta(d)\|_{H}\lesssim r_E(\delta,n^{\EXP},b^{\EXP},c^{\EXP})+r_{\mu}(\delta,n)$. By Assumption~\ref{assumption:smooth}, $\|\gamma_0^{\OBS}\|_H$ is bounded by a constant. By Assumption~\ref{assumption:RKHS}, $\left\| \int \{\mu_s^{\EXP}(d,x)\otimes \phi(x)\}\d\P(x)\right\|_H \leq \kappa_s\kappa_x$. 

In summary, for $n^{\OBS}$ and $n^{\EXP}$ sufficiently large, with probability $1-3\delta$,
$$
\sup_{d\in \cD} |\hat{\theta}(d)-\theta_0(d)|\lesssim r(\delta,n^{\OBS},b^{\OBS},c^{\OBS})+r_E(\delta,n^{\EXP},b^{\EXP},c^{\EXP})+r_{\mu}(\delta,n).
$$
Substituting in these rates from Lemmas~\ref{lemma:regression},~\ref{lemma:unconditional}, and~\ref{lemma:conditional} yields the desired result for $\theta_0(d)$. In particular, $r_{\mu}(\delta,n)$ is always smaller than $r(\delta,n^{\OBS},b^{\OBS},c^{\OBS})+r_E(\delta,n^{\EXP},b^{\EXP},c^{\EXP})$. The arguments for the other long term dose response curves are similar.
\end{proof}

\begin{proof}[Proof of Theorem~\ref{theorem:consistency_dist}]
     Fix $d\in\cD$. By Theorem~\ref{theorem:representation_dist}, 
    \begin{align*}
        \hat{\nu}(d)-\nu_0(d)
        &=(\hat{E}^{\OBS})^* \left[\frac{1}{n}\sum_{i=1}^n \{\hat{\mu}_s^{\EXP}(d,X_i)\otimes \phi(X_i)\}\right]
        - (E_0^{\OBS})^*\left[ \int \{\mu_s^{\EXP}(d,x)\otimes \phi(x)\}\d\P(x)\right]\\
        &=(\hat{E}^{\OBS})^*\{\Delta(d)\}
        + (\hat{E}^{\OBS}-E_0^{\OBS})^* \left[\int \{\mu_s^{\EXP}(d,x)\otimes \phi(x)\}\d\P(x)\right] \\
        &=(\hat{E}^{\OBS}-E_0^{\OBS})^*\{\Delta(d)\}
        +(E_0^{\OBS})^*\{\Delta(d)\}
        + (\hat{E}^{\OBS}-E_0^{\OBS})^* \left[ \int \{\mu_s^{\EXP}(d,x)\otimes \phi(x)\}\d\P(x)\right].
    \end{align*}
    Therefore, by the triangle inequality, the definition of the operator norm, and the fact that the Hilbert-Schmidt norm upper bounds the operator norm, $\|\hat{\nu}(d)-\nu_0(d)\|_{H_{\cY}}$ is bounded by
\begin{align*}
    & \left\|(\hat{E}^{\OBS}-E_0^{\OBS})^*\{\Delta(d)\}\right\|_{H_{\cY}}
        +\left\|(E_0^{\OBS})^*\{\Delta(d)\}\right\|_{H_{\cY}}
        + \left\|(\hat{E}^{\OBS}-E_0^{\OBS})^* \left[ \int \{\mu_s^{\EXP}(d,x)\otimes \phi(x)\}\d\P(x)\right]\right\|_{H_{\cY}} \\
        &\leq \|\hat{E}^{\OBS}-E_0^{\OBS}\|_{\op} \left\|\Delta(d)\right\|_{H}
        +\|E_0^{\OBS}\|_{\op}\left\|\Delta(d)\right\|_{H}
        + \|\hat{E}^{\OBS}-E_0^{\OBS}\|_{\op}\left\| \int \{\mu_s^{\EXP}(d,x)\otimes \phi(x)\}\d\P(x)\right\|_{H} \\
        &\leq \|\hat{E}^{\OBS}-E_0^{\OBS}\|_{\cL_2} \left\|\Delta(d)\right\|_{H}
        +\|E_0^{\OBS}\|_{\cL_2}\left\|\Delta(d)\right\|_{H}
        + \|\hat{E}^{\OBS}-E_0^{\OBS}\|_{\cL_2}\left\| \int \{\mu_s^{\EXP}(d,x)\otimes \phi(x)\}\d\P(x)\right\|_{H},
\end{align*}
where in the final line we abbreviate $\cL_2=\cL_2(H_{\cY},H_{\cS}\otimes H_{\cX})$.

By Lemma~\ref{lemma:conditional}, for $n^{\OBS}$ sufficiently large, with probability $1-\delta$,  $\|\hat{E}^{\OBS}-E_0^{\OBS}\|_{\cL_2} \lesssim r(\delta,n^{\OBS},b^{\OBS},c^{\OBS}).$ By Proposition~\ref{prop:rate}, for $n^{\EXP}$ sufficiently large, with probability $1-2\delta$, $\sup_{d\in\cD}\|\Delta(d)\|_{H}\lesssim r_E(\delta,n^{\EXP},b^{\EXP},c^{\EXP})+r_{\mu}(\delta,n)$. By Assumption~\ref{assumption:smooth_op_dist}, $\|E_0^{\OBS}\|_{\cL_2}$ is bounded by a constant. By Assumption~\ref{assumption:RKHS}, $\left\| \int \{\mu_s^{\EXP}(d,x)\otimes \phi(x)\}\d\P(x)\right\|_H \leq \kappa_s\kappa_x$. 

In summary, for $n^{\OBS}$ and $n^{\EXP}$ sufficiently large, with probability $1-3\delta$,
$$
\sup_{d\in \cD}\|\hat{\nu}(d)-\nu_0(d)\|_{H_{\cY}} \lesssim r_E(\delta,n^{\OBS},b^{\OBS},c^{\OBS})+r_E(\delta,n^{\EXP},b^{\EXP},c^{\EXP})+r_{\mu}(\delta,n).
$$
Substituting in these rates from Lemmas~\ref{lemma:unconditional} and~\ref{lemma:conditional} yields the desired result for $\nu_0(d)$. In particular, $r_{\mu}(\delta,n)$ is always smaller than $r_E(\delta,n^{\OBS},b^{\OBS},c^{\OBS})+r_E(\delta,n^{\EXP},b^{\EXP},c^{\EXP})$. The arguments for the other long term counterfactual embeddings are similar.
\end{proof}

\begin{proof}[Proof of Corollary~\ref{cor:dist}]
By Theorem~\ref{theorem:consistency_dist}, for any $d\in\cD$, 
    $$
    \|\hat{\nu}(d)-\nu_0(d)\|_{H_{\cY}}=O_p\left\{(n^{\OBS})^{-\frac{1}{2}\frac{c^{\OBS}-1}{c^{\OBS}+1/b^{\OBS}}}+ (n^{\EXP})^{-\frac{1}{2}\frac{c^{\EXP}-1}{c^{\EXP}+1/b^{\EXP}}} \right\}.
    $$
    Denote the samples constructed by Algorithm~\ref{algo:distribution} by $(Y_j')$  where $j=1,...,m$. Then by \cite[Section 4.2]{bach2012equivalence},
    $
    \left\|\hat{\nu}(d)-\frac{1}{m}\sum_{j=1}^m \phi(Y_j')\right\|_{H_{\cY}}=O(m^{-1/2}).
    $
    Therefore by the triangle inequality,
    $$
    \left\|\frac{1}{m}\sum_{j=1}^m \phi(Y_j')-\nu_0(d)\right\|_{H_{\cY}}=O_p\left\{(n^{\OBS})^{-\frac{1}{2}\frac{c^{\OBS}-1}{c^{\OBS}+1/b^{\OBS}}}+ (n^{\EXP})^{-\frac{1}{2}\frac{c^{\EXP}-1}{c^{\EXP}+1/b^{\EXP}}} +m^{-1/2} \right\}.
    $$
    The desired result follows from \cite{sriperumbudur2016optimal}, as quoted by \cite[Theorem 1.1]{simon2020metrizing}. The arguments for other long term counterfactual distributions are identical.
\end{proof}

\subsection{Connection to semiparametric inference}

An earlier draft \cite{singh2022generalized} further demonstrated how Theorem~\ref{theorem:consistency} satisfies the rate conditions for inference on the effect of a binary action.  In particular, it proved nonasymptotic Gaussian approximation  by verifying the conditions of \cite{meza2021nested}, when the action is binary. This draft focuses on the continuous action setting, which has received less attention in long term causal inference.
\section{Application details}\label{sec:experiment_details}

\subsection{Sample and variable definitions}

We use the same sample as \cite{athey2020combining}, where $n^{\EXP}=2,052$ from Project STAR and $n^{\OBS}=10,240$ from NYC public schools. In particular, the NYC students are from cohorts with year of birth from $1985$ to $1995$. For the benchmark that uses covariates, we keep only complete observations, yielding $n^{\OBS}=10,222$.  

The test score definitions follow \cite{krueger2001effect,chetty2011does,athey2020combining}. They combine both math and reading scores on the grade-appropriate Stanford Achievement Test, which is multiple choice.

The covariates are binary: race indicates whether a student is white, gender indicates whether a student is female, and free and reduced price lunch status indicates whether a student is eligible for discounted meals at school.

\subsection{Implementation details}

We use Gaussian kernels for class size and test scores. Following Appendix~\ref{sec:tuning}, we set lengthscales according to the median interpoint distance heuristic. When using covariates, we use a binary kernel for each covariate. We combine multiple kernels by taking the product.

We tune the regularization hyperparameters using leave-one-out cross validation as described in Appendix~\ref{sec:tuning}. We set $\lambda^{\OBS}$ according to Algorithm~\ref{algo:cv} and $\lambda^{\EXP}$ according to Algorithm~\ref{algo:cv_op}. 

We follow similar tuning procedures for the oracle and benchmarks, which are versions of previous work \cite{singh2020kernel}. In particular, we set the regularization according to Algorithm~\ref{algo:cv}, as suggested by those authors.

\subsection{Alternative samples}

\begin{figure}
\captionsetup[subfigure]{justification=Centering}
\begin{subfigure}[t]{0.32\textwidth}
         \centering
        \resizebox{\textwidth}{!}{%
       \includegraphics[width=\textwidth]{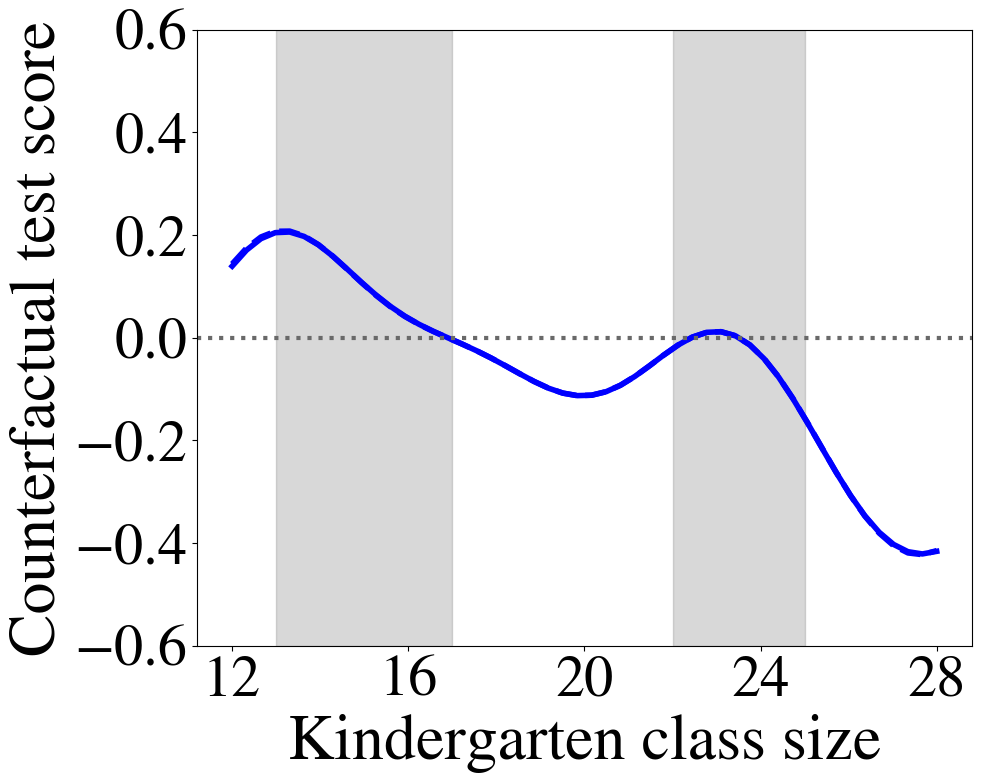}
        }
    \caption{Third grade.}\label{fig:prop3}
\end{subfigure}\hspace{\fill} 
\begin{subfigure}[t]{0.32\textwidth}
          \centering
        \resizebox{\textwidth}{!}{%
      \includegraphics[width=\textwidth]{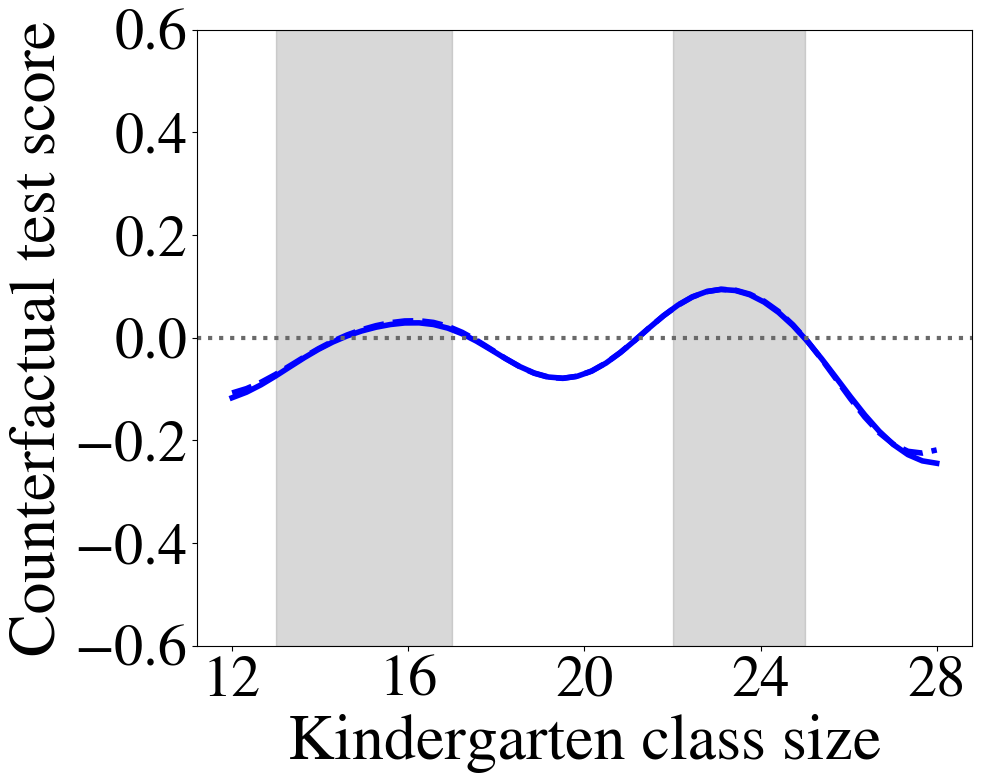}
        }
    \caption{Fourth grade.}\label{fig:prop4}
\end{subfigure}
\hspace{\fill} 
\begin{subfigure}[t]{0.32\textwidth}
          \centering
        \resizebox{\textwidth}{!}{%
      \includegraphics[width=\textwidth]{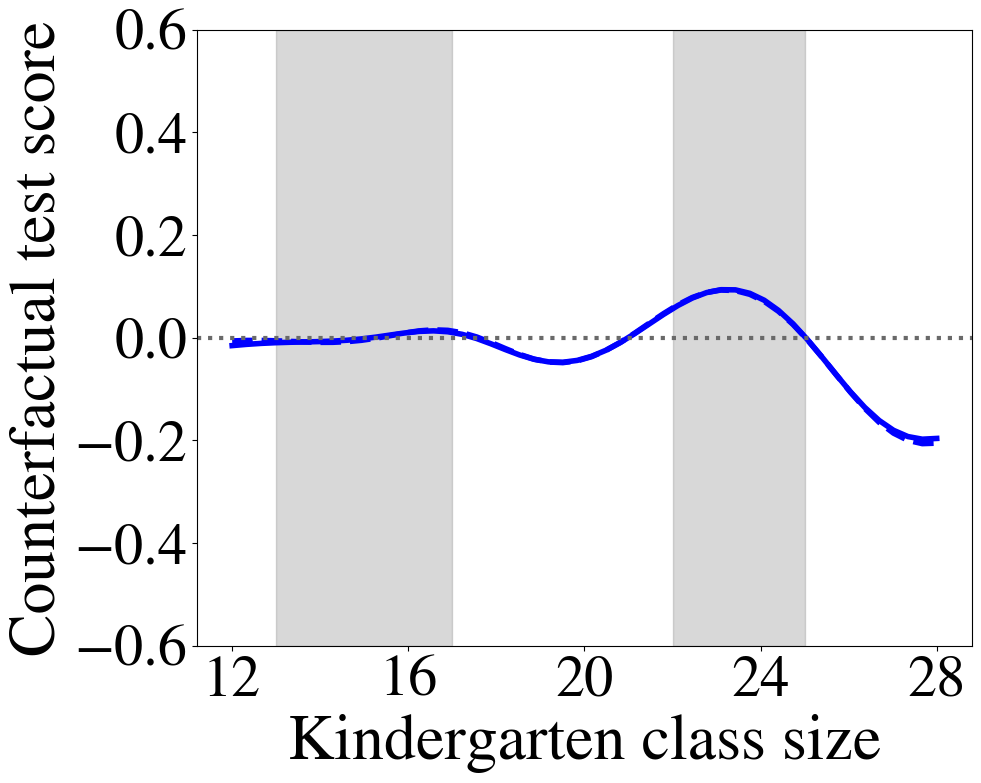}
        }
    \caption{Fifth grade.}\label{fig:prop5}
\end{subfigure}
\caption{Our method is robust to the choice of sample. 
We compare our proposal for the missing-at-random model, either with the full sample (dark blue, solid) or a subsetted sample (dark blue, dashed). In the background, we indicate ranges of class sizes in the Project STAR protocol (gray).}\label{fig:345_prop}
\end{figure}

We find that choosing a different sample of NYC students does not change the results. In particular, we compare our missing-at-random proposal using the sample from the main text ($n^{\OBS}=10,240$) with our missing-at-random proposal using a subsetted sample keeping only class sizes in $[12,28]$ ($n^{\OBS}=7,477$). Figure~\ref{fig:345_prop} confirms that the estimates are almost identical.

\subsection{Longer horizons}

Finally, we repeat the exercise in Section~\ref{sec:application} over longer horizons. We document the same phenomena as in the main text. The gap between our proposals and the benchmarks gradually becomes smaller as the horizon becomes longer.

\begin{figure}
\captionsetup[subfigure]{justification=Centering}
\begin{subfigure}[t]{0.32\textwidth}
         \centering
        \resizebox{\textwidth}{!}{%
       \includegraphics[width=\textwidth]{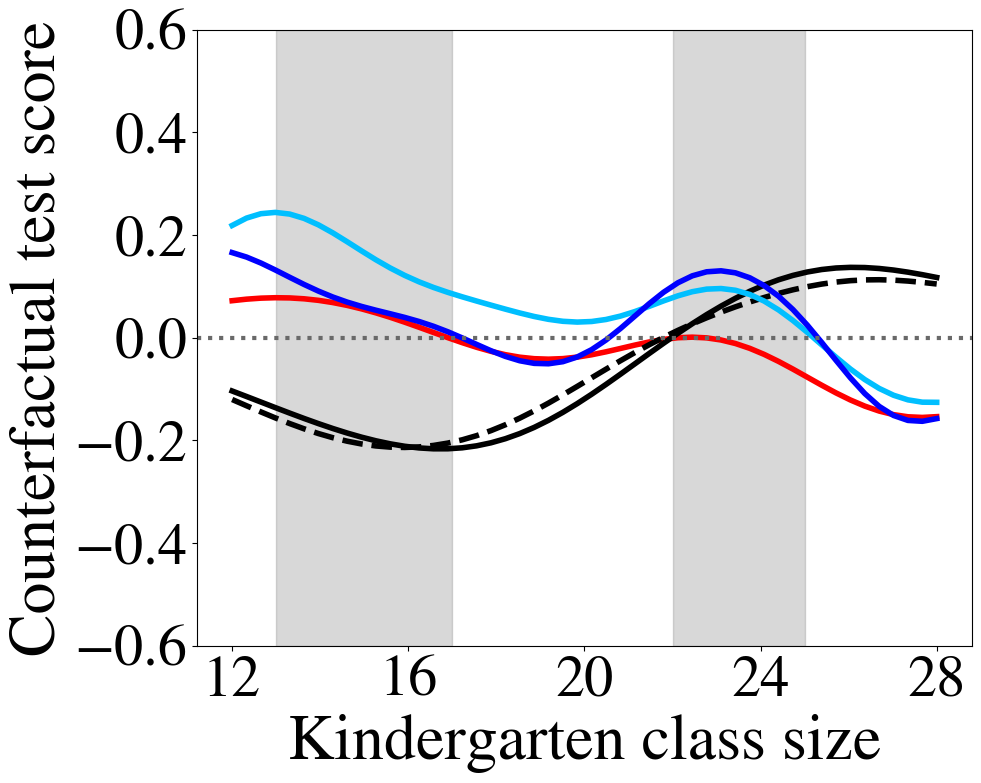}
        }
    \caption{Sixth grade.}\label{fig:semi6_grey}
\end{subfigure}\hspace{\fill} 
\begin{subfigure}[t]{0.32\textwidth}
          \centering
        \resizebox{\textwidth}{!}{%
      \includegraphics[width=\textwidth]{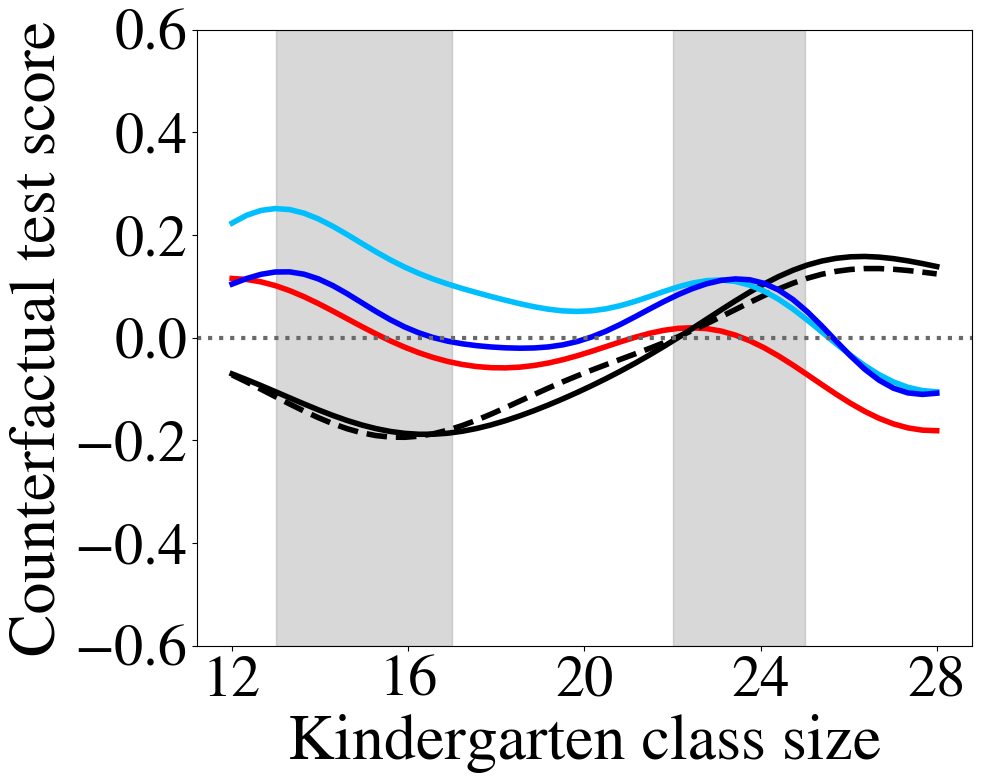}
        }
    \caption{Seventh grade.}\label{fig:semi7_grey}
\end{subfigure}
\hspace{\fill} 
\begin{subfigure}[t]{0.32\textwidth}
          \centering
        \resizebox{\textwidth}{!}{%
      \includegraphics[width=\textwidth]{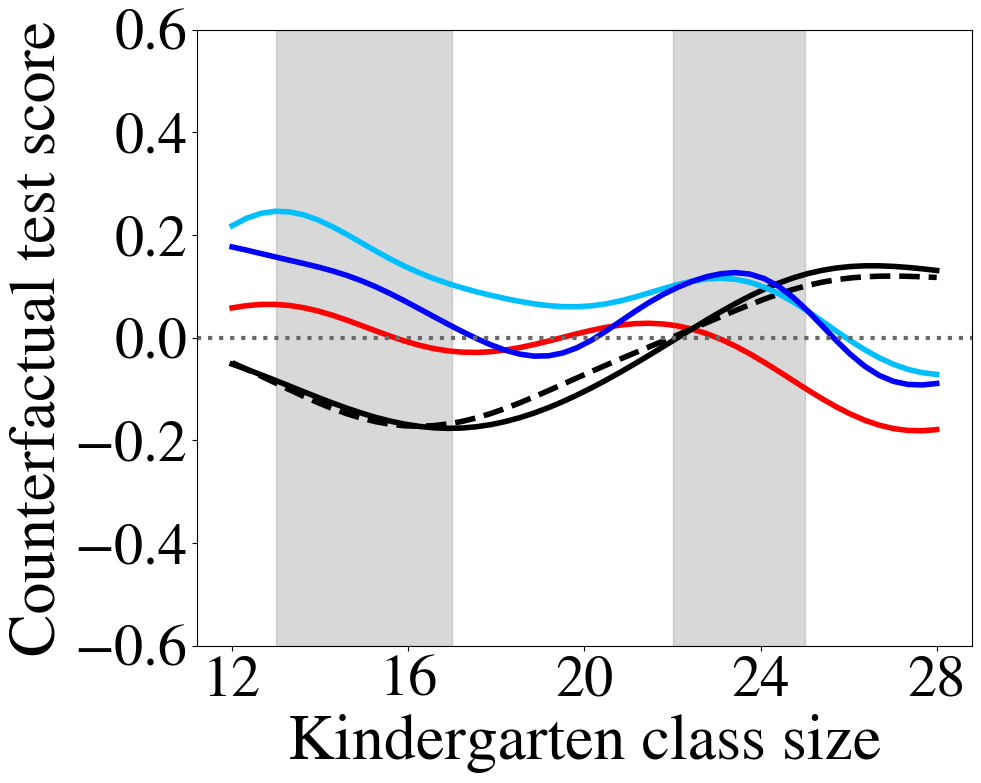}
        }
    \caption{Eighth grade.}\label{fig:semi8_grey}
\end{subfigure}
\caption{Our method outperforms some benchmarks for the long term dose response of Project STAR over longer horizons. 
We compare an oracle (red) with our proposals for the surrogate model (light blue) and the missing-at-random model (dark blue). 
We also visualize benchmarks from previous work, either with (black, dashed) or without (black, solid) covariates. In the background, we indicate ranges of class sizes in the Project STAR protocol (gray).}\label{fig:678_grey}
\end{figure}

\end{document}